\documentclass[12pt]{iopart}

\usepackage{qcircuit}
\usepackage{graphicx}
\usepackage{braket} 
\usepackage{siunitx} 
\usepackage{subcaption}
\usepackage{pdfpages}

\expandafter\let\csname equation*\endcsname\relax
\expandafter\let\csname endequation*\endcsname\relax
\usepackage{amsmath}

\begin{document}

\title{Entanglement filter with Rydberg atoms}

\author{Gen-Sheng Ye$^{1\ast}$, Biao Xu$^{1\ast}$, Yue Chang$^{2,3\ast}$, Shuai Shi$^{1}$, Tao Shi$^{4,5}$, Lin Li$^{1\dagger}$}
\address{$^1$MOE Key Laboratory of Fundamental Physical Quantities Measurement,\\
Hubei Key Laboratory of Gravitation and Quantum Physics, PGMF,\\
Institute for Quantum Science and Engineering, School of Physics,\\
Huazhong University of Science and Technology, Wuhan 430074, China}

\address{$^2$Beijing Automation Control Equipment Institute, Beijing 100074, China}
\address{$^3$Quantum Technology R$\&$D Center of China Aerospace Science and Industry Corporation, Beijing 10019, China}

\address{$^4$Institute of Theoretical Physics, Chinese Academy of Sciences, P.O. Box 2735, Beijing 100190, China}
\address{$^5$CAS Center for Excellence in Topological Quantum Computation, University of Chinese Academy of Sciences, Beijing 100049, China}

\address{$\ast$These authors contributed equally.}
\address{$\dagger$ To whom correspondence should be addressed; E-mail: li\_lin@hust.edu.cn}

\vspace{10pt}
\begin{indented}
\item[]\date{\today}
\end{indented}

\begin{abstract}
\textbf{Devices capable of deterministically manipulating the photonic entanglement are of paramount importance, since photons are the ideal messengers for quantum information.
Here, we report a Rydberg-atom-based entanglement filter that preserves the desired photonic entangled state and deterministically blocks the transmission of the unwanted ones.
Photonic entanglement with near-unity fidelity can be extracted from an input state with an arbitrarily low initial fidelity.
The protocol is inherently robust, and succeeds both in the Rydberg blockade regime and in the interaction-induced dissipation regime.
Such an entanglement filter opens new routes toward scalable photonic quantum information processing with multiple ensembles of Rydberg atoms.}
\end{abstract}

Advancing the efficient quantum control of photonic entanglement is at the heart of quantum science~({\it 1–4\/}).
As one of the key elements in quantum photonics, an entanglement filter (EF) transmits the entanglement of the desired quantum states, while blocking the transmission of unwanted photonic components~({\it 5\/}).
It has a plethora of potential applications, including photonic entanglement generation~({\it 6\/}), 
all-optical quantum information processing~({\it 7,8\/}) and entanglement distillations~({\it 9,10\/}).
However, its scalability and applicability have been limited by the fact that all photonic entanglement filter protocols to date are based on linear-optical approaches~({\it 5,11–13\/}), which remove unwanted photonic states only in a probabilistic way. The probabilistic nature and the requirement of ancillary quantum resources lead to poor scalability and overwhelming resource consumption. Moreover, the output entanglement fidelity in the linear-optical approach is ultimately limited by the finite interference visibility between photons.

Therefore, scalable quantum photonic applications~({\it 1–3,14\/}) will largely benefit from an ideal entanglement filter that deterministically removes undesired states, unconditionally achieves a high entanglement fidelity, and demands no extra photonic resources. Unfortunately, the realization of such a superior entanglement-filter protocol remains elusive \textit{hitherto}, due to the lack of a strong and controllable 
photon-photon interaction in linear-optical approaches.
In recent years, cold Rydberg atoms have been employed to facilitate the interactions between photons
and to achieve intrinsically deterministic quantum photonic operations, such as single-photon generation~({\it 15–17\/}) and manipulation~({\it 18–22\/}), atom-photon entanglement preparation~({\it 23,24\/}), and photon-photon gate~({\it 25\/}).

Here, we propose and demonstrate an above-envisioned superior entanglement filter by exploiting cold Rydberg atomic ensembles to mediate the interaction between photons.
The qubits are encoded in the horizontal (H) and vertical (V) polarizations of photons \textit{a} and \textit{b}.
With a product state of $(\ket{H}_a + \ket{V}_a)(\ket{H}_b + \ket{V}_b)$ as input, the filter removes the $\ket{H}_a\ket{H}_b$ and $\ket{V}_a\ket{V}_b$ components, resulting in a maximally entangled Bell state $\ket{\Psi^+} =1/\sqrt{2} \left( \ket{H}_a\ket{V}_b + \ket{V}_a\ket{H}_b \right)$.
In contrast to previous probabilistic linear-optical approaches, our protocol eliminates the unwanted states in a deterministic way, outputs high-fidelity entangled states, and requires no extra quantum resources.


\begin{figure*}
  \centering
  \includegraphics[width=\textwidth]{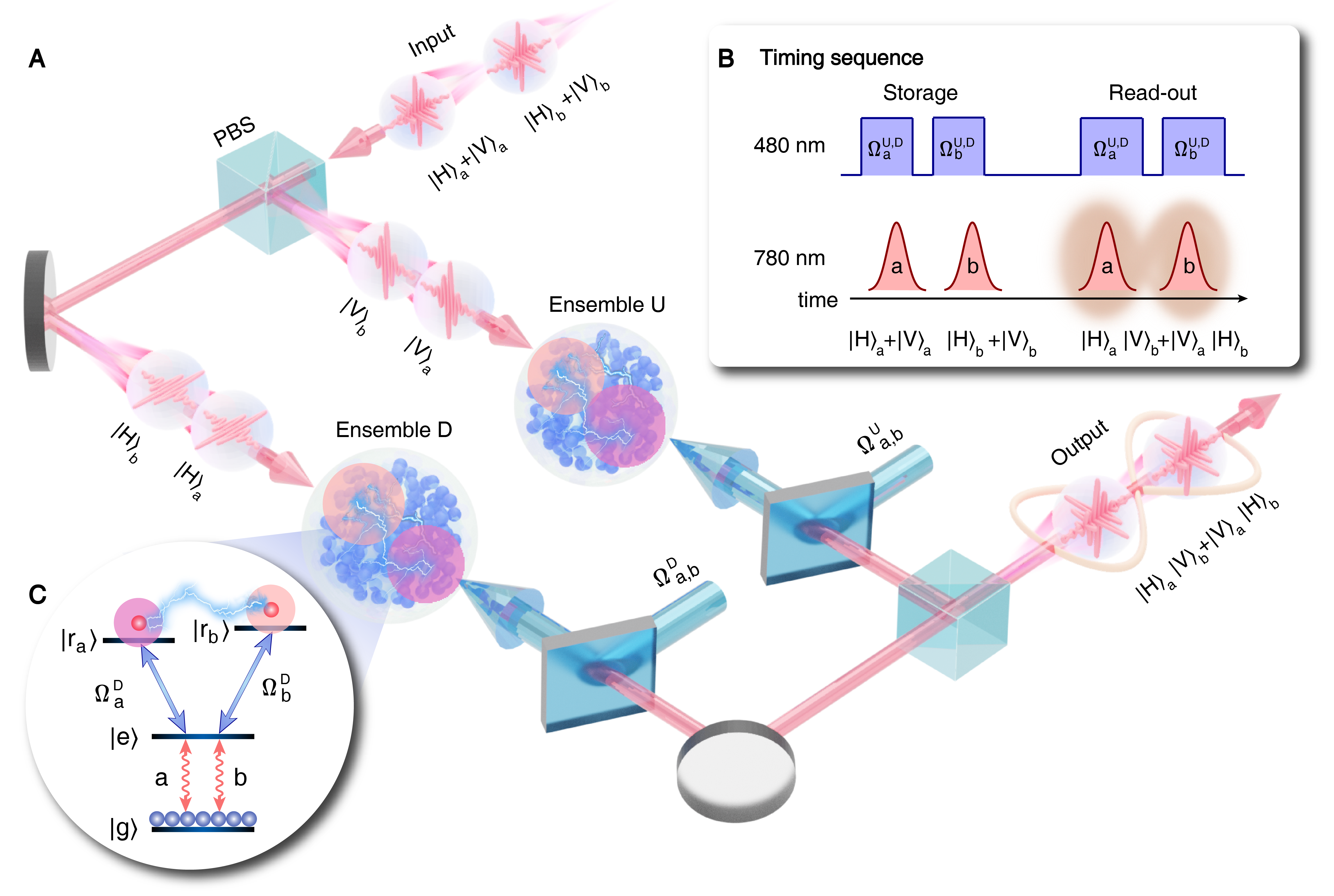}
  \caption{
	\textbf{Illustration of experimental protocol.}
     \textbf{(A)}
     The horizontally (H) and vertically (V) polarized components of the \SI{780}{\nano m} photonic qubits $a$ and $b$ are separated into the down (D) and upper (U) paths by a polarization beam splitter (PBS), and  stored as Rydberg excitations by the \SI{480}{\nano m} laser fields $\Omega_{a}^{\mathrm{U},\mathrm{D}}$ and $\Omega_{b}^{\mathrm{U},\mathrm{D}}$. 
     The \SI{780}{\nano m} and \SI{480}{\nano m} fields are counter-propagating and focused on the cold $^{87}$Rb atomic ensembles with waists of \SI{6}{\micro m} and \SI{15}{\micro m}, respectively.
     In the read-out stage, the \SI{480}{\nano m} laser fields $\Omega_{a}^{\mathrm{U},\mathrm{D}}$ and $\Omega_{b}^{\mathrm{U},\mathrm{D}}$ are sequentially applied to convert the Rydberg excitations $\ket{D}_a / \ket{U}_a$ and $\ket{D}_b / \ket{U}_b$ to photonic qubits $\ket{H}_a / \ket{V}_a$ and $\ket{H}_b / \ket{V}_b$. 
     The photons then go through a polarization-projective measurement setup before being collected by single-mode fibers and detected by SPCMs.
     The inserts show the timing sequence \textbf{(B)} and the relevant $^{87}$Rb atomic levels \textbf{(C)}: ground state $\ket{\mathrm{g}}=\ket{5S_{1/2},F=2,m_F=2}$, intermediate state $\ket{\mathrm{e}}=\ket{5P_{3/2},F=3,m_F=3}$, and Rydberg states $\ket{r_a}=\ket{n_aD_{5/2},J=5/2,m_j=5/2}$ and $\ket{r_b}=\ket{n_bD_{5/2},J=5/2,m_j=5/2}$.
     }
  \label{fig1}
\end{figure*}

As illustrated in Fig.~1, our experiment employs two Rydberg ensembles to achieve polarization-selective interaction between photonic qubits.
The working principle is to convert the unwanted photonic components into double Rydberg excitations in the same atomic ensemble and to achieve the deterministic removal of these states with either the Rydberg blockade~({\it 26\/}) or the interaction induced two-body dissipation~({\it 27\/}).
Using the photon storage based on the Rydberg electromagnetically induced transparency (EIT), the \SI{480}{\nano\meter} control laser field $\Omega_a^\mathrm{U} (\Omega_a^\mathrm{D})$ coherently transfers the \SI{780}{\nano\meter} photon in the state $\ket{H}_a$ ($\ket{V}_a$) into a collective atomic excitation in the down (upper) ensemble:

\begin{small}
\begin{equation}
\begin{aligned}
 &\ket{D}_a=
 \sum^N_{j=1}{\ket{\mathrm{g^D}}_1\ldots
 \ket{\mathrm{{r_a}^D}}_j\ldots
 \ket{\mathrm{g^D}}_N
 }/\sqrt{N}
  \\[0mm]
 &\ket{U}_a=
 \sum^N_{j=1}{\ket{\mathrm{g^U}}_1\ldots
 \ket{\mathrm{{r_a}^U}}_j\ldots
 \ket{\mathrm{g^U}}_N
 }/\sqrt{N} ,
\end{aligned}
\end{equation}
\end{small}where $\ket{\mathrm{g}}$ is the atomic ground state, $\ket{\mathrm{r}_a}$ is a Rydberg state, and U/D denotes the upper/down ensemble. 
To induce the interaction between photons \textit{a} and \textit{b}, an adjacent Rydberg level $\ket{\mathrm{r}_b}$ is employed to transfer the qubit state $\ket{H}_b$ ($\ket{V}_b$) to the excitation  $\ket{D}_b$ ($\ket{U}_b$) in the down (upper) ensemble.

To implement the entanglement filter in the blockade regime, high-lying Rydberg states with principal quantum numbers $n_a = 76$ and $n_b = 77$  are employed.
As a result, the storage of unwanted components $\ket{H}_a\ket{H}_b$ and $\ket{V}_a\ket{V}_b$ into double Rydberg excitations in the same ensemble, i.e., $\ket{D}_a\ket{D}_b$ or $\ket{U}_a\ket{U}_b$, is strongly suppressed by the blockade effect.
On the other hand, two ensembles are separated by \SI{150}{\micro\meter}, well beyond the Rydberg blockade radius, such that the desired entangled state $\ket{\Psi^+}$ can be stored as $1/\sqrt{2} \left (\ket{D}_a\ket{U}_b + \ket{U}_a\ket{D}_b \right)$ of the atoms.
After the storage process, the Rydberg excitations are converted back to the photonic qubits in the state $\ket{\Psi^+}$ through collective emission by sequentially applying \SI{480}{\nano\meter} read-out light fields.

\begin{figure*}[ht]
  \centering
  \includegraphics[width=\textwidth]{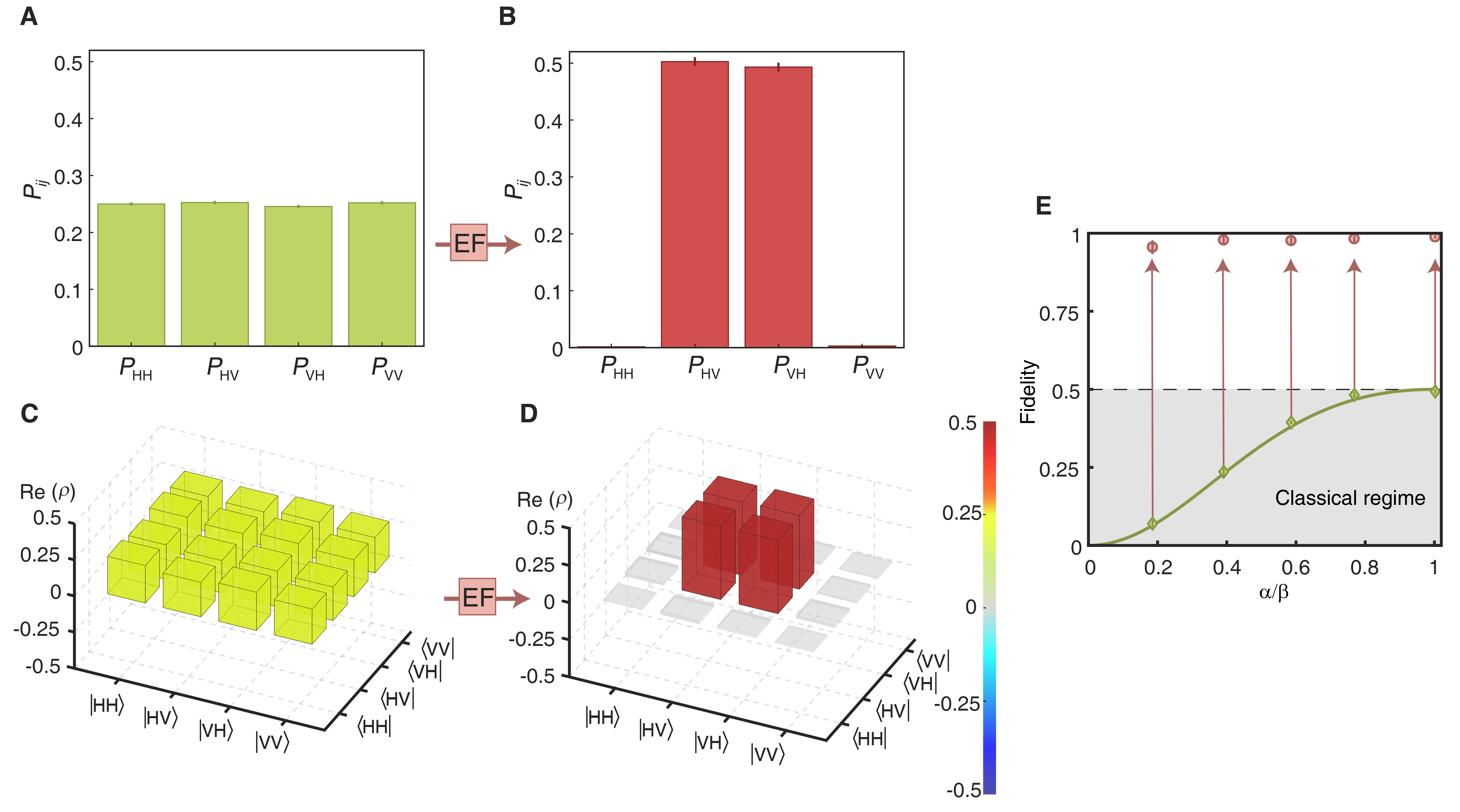}
  \caption{
	\textbf{Blockade-based entanglement filter.}
	\textbf{(A and B) }
	 Normalized two-photon populations of the input (A) and output (B) state, obtained from photon correlation measurements. 
     \textbf{(C and D) } The real part of the reconstructed density matrix of the input (C) and the output (D) state.
     \textbf{(E) }The measured input- (green diamonds) and output- (red circles) state fidelities as functions of the input-state amplitude ratio $\alpha/\beta$. Solid line: expected input fidelity at different $\alpha/\beta$. Error bars: $1\sigma$ standard deviation from photoelectric counting events.
    }
  \label{Fig:2}
\end{figure*}

\begin{figure}
  \centering
  \includegraphics[width=\columnwidth]{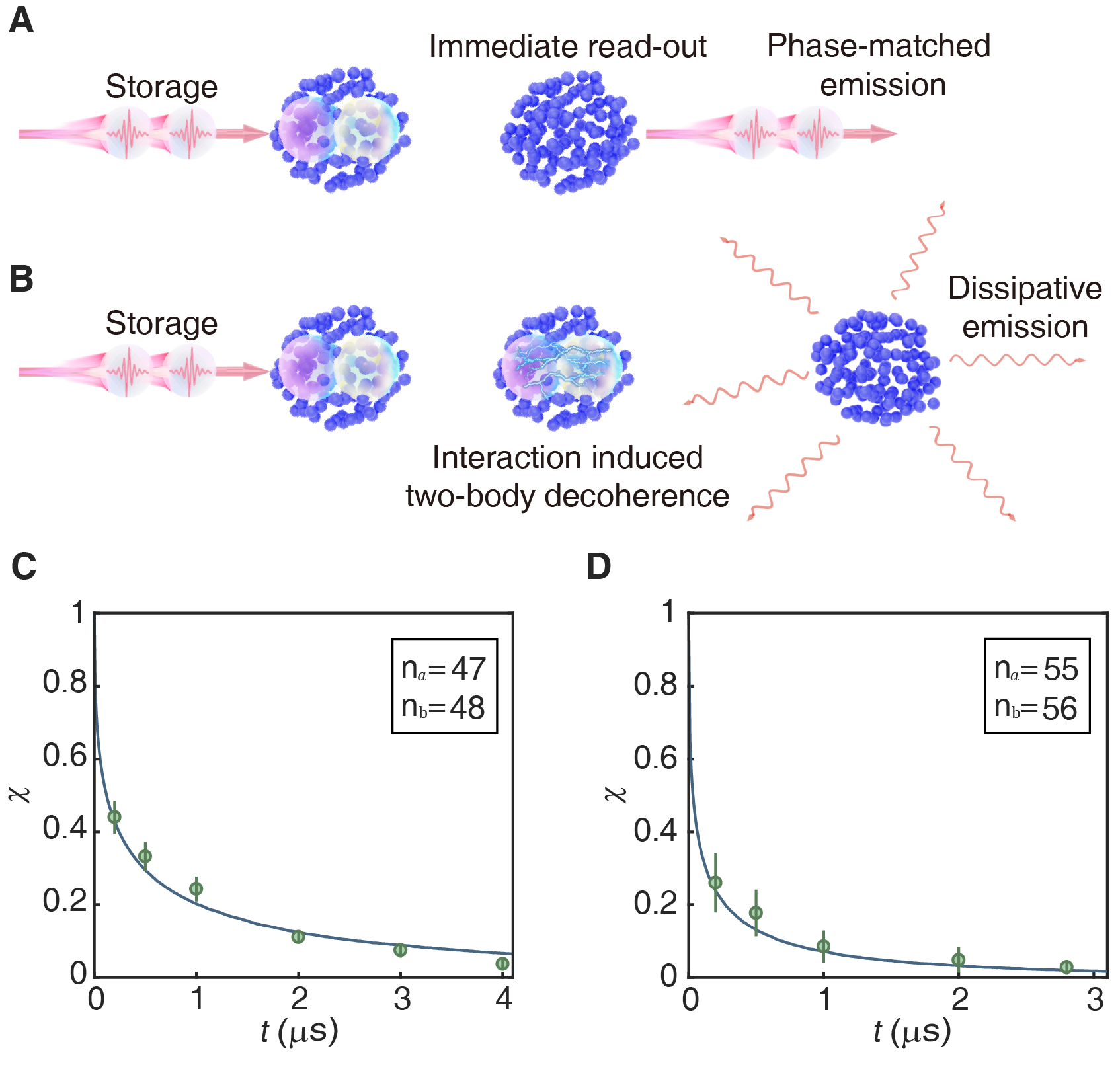}
  \caption{
	\textbf{Interaction induced two-body decoherence.}
      \textbf{(A and B)} Illustration of two-body decoherence induced by distance-dependent interaction.
      \textbf{(C and D)} The suppression ratio $\chi=\frac{P_{HH}+P_{VV}}{P_{HV}+P_{VH}}$ as a function of evolution time $t$ for $n_a=47, n_b=48$ (C) and $n_a=55, n_b=56$ (D). Error bars: $1\sigma$ standard deviation from photoelectric counting events. Solid curves: simulations based on the two-body decoherence mechanism.
   }
  \label{Fig:3}
\end{figure}

The polarization-selective photon blockade effect is shown in Fig.~2, A and B.
Two weak coherent laser pulses with average photon number $\braket{n} \sim 0.1$ are used to approximate single photons \textit{a} and \textit{b}.
After passing through the entanglement filter, photons are detected in the $\ket{H}/\ket{V}$ basis by single-photon counting modules (SPCMs) . 
Figure~2A displays the measured two-photon populations of the input state, which are distributed equally ($\sim 0.25$) over all bases.
The populations of the output state are shown in Fig.~2B, where the $\ket{H}_a\ket{H}_b$ and $\ket{V}_a\ket{V}_b$ components are strongly suppressed. 
The measured suppression ratio $\chi=\frac{P_{HH}+P_{VV}}{P_{HV}+P_{VH}}\sim \SI{4(1)}{\times 10^{-3}}$ demonstrates the strong blockade effect in the entanglement filter.

To characterize the entanglement filter in the non-classical regime, quantum state tomography for the input and output states are performed.
The density matrix $\rho_\mathrm{in}$ of the input state is reconstructed and shown in Fig.~2C, which has a fidelity of $F_\mathrm{in}=49.3(7)\%$ overlapping with the target state $\ket{\Psi^+}$. The entanglement filter removes the undesired components and improves the fidelity up to $F_\mathrm{out}=98.8(5)\%$ (Fig.~2D).
The reduction of $F_\mathrm{out}$ from unity is mainly caused by the $0.61(6)\%$ infidelity from background detection events of SPCMs, and the background-corrected fidelity is improved to $F_\mathrm{cor}=99.4(5)\%$.
The multi-photon components from the weak coherent light and the crosstalk between different polarization bases also cause infidelities of $\le0.58\%$ and $0.25(23)\%$, respectively.
The above-mentioned sources of infidelities are not intrinsic to our protocol and can be further suppressed with technical efforts.
The fundamental limitation on the fidelity comes from the imperfect suppression of double Rydberg excitations, which gives an upper bound of $F_{out}\sim 99.98(6)\%$ (see supplementary materials).

An ideal entanglement filter should post no prerequisite on the fidelity of the input state and extract the desired entangled state from an arbitrarily noisy input.
To demonstrate this essential capability, ($\alpha \ket{H}_a+\beta \ket{V}_a$)($\alpha \ket{H}_b+\beta \ket{V}_b$) is employed as the input state.
By varying the amplitude ratio $\alpha / \beta$, the input fidelity $F_{in}$ can be tuned in the classical regime between 0 and 0.5 (green diamonds in Fig.~\ref{Fig:2}E).
When the entanglement filter is applied, the fidelities of the corresponding output states are improved to a near-unity level (red circles in Fig.~\ref{Fig:2}E).
In the case of $\alpha/\beta = 0.185$, our entanglement filter improves the input state fidelity $F_{in}=0.070(12)$ by more than one order of magnitude.
The measured state fidelity $F_{out}=0.955(22)$ is limited by the decreased signal-to-background ratio for small $\alpha / \beta$, and the background-corrected fidelity $F_{cor}=0.992(25)$ is not affected by the low input state fidelity.

Developing novel Rydberg quantum photonic protocols with low principal quantum number $n$ holds the promise to alleviate some of the decoherence and loss mechanisms associated with high $n$ states, such as long-lived Rydberg contaminants, energy level drifts induced by background electric fields, and density-dependent dephasing.
To this end, we demonstrate an entanglement filter in the absence of blockade effect using lower-lying Rydberg states $n_a = 47$ and $n_b = 48$, whose van der Waals interaction coefficient $C_6$ is nearly two orders of magnitudes weaker than that of the states $n_a = 76$ and $n_b = 77$.
Therefore, the $\ket{H}_a\ket{H}_b$ and $\ket{V}_a\ket{V}_b$ components can be stored as double Rydberg excitations in the same ensemble.

The distance-dependent van der Waals interaction strength $V_{jj^\prime}=\frac{C_{6}}{R_{jj^\prime}^6}$ varies strongly for Rydberg atom pairs with different separations $R_{jj^\prime}$, and leads to the accumulation of random phases and a fast two-body decoherence during the quantum evolution~({\it 27\/}):

\begin{small}
\begin{equation}
 \ket{U}_a\ket{U}_b\propto
  \sum^N_{j,j^\prime \neq j}{
  e^{i\frac{V_{jj^\prime}}{\hbar}t}
  \ket{\mathrm{g^U}}_1\ldots
  \ket{\mathrm{{r_a}^U}}_j\ldots\ket{\mathrm{{r_b}^U}}_{j^\prime}\ldots
  \ket{\mathrm{g^U}}_N
.}
\end{equation}
\end{small}

If the read-out is performed immediately after storage, i.e., without two-body decoherence, the spatial mode of the retrieved photons will be highly directional, as a result of the mode-matched collective emission (Fig.~3A).
However, the accumulation of interaction-induced random phases deteriorates the collective coherence of the excitations and leads to the spontaneous photon emission into random directions during the read-out (Fig.~3B).
In the experiment, only photons in the mode-matched direction are collected and therefore the retrieved $\ket{H}_a\ket{H}_b$ and $\ket{V}_a\ket{V}_b$ components are suppressed due to the dissipation.
The entangled state $\ket{\Psi^+}$ is stored in the decoherence-free-subspace $1/\sqrt{2} \left (\ket{D}_a\ket{U}_b + \ket{U}_a\ket{D}_b \right)$, hence immune to the above dissipation. Figure~3C displays the measured suppression ratio $\chi$ as a function of the dissipative evolution time $t$. The ratio $\chi$ decreases to $0.440(45)$ with just a short evolution time of $\SI{0.2}{\mu\second}$ and is further suppressed to $0.037(17)$ after $\SI{4}{\mu\second}$.

\begin{figure*}[ht]
  \centering
  \includegraphics[width=\textwidth]{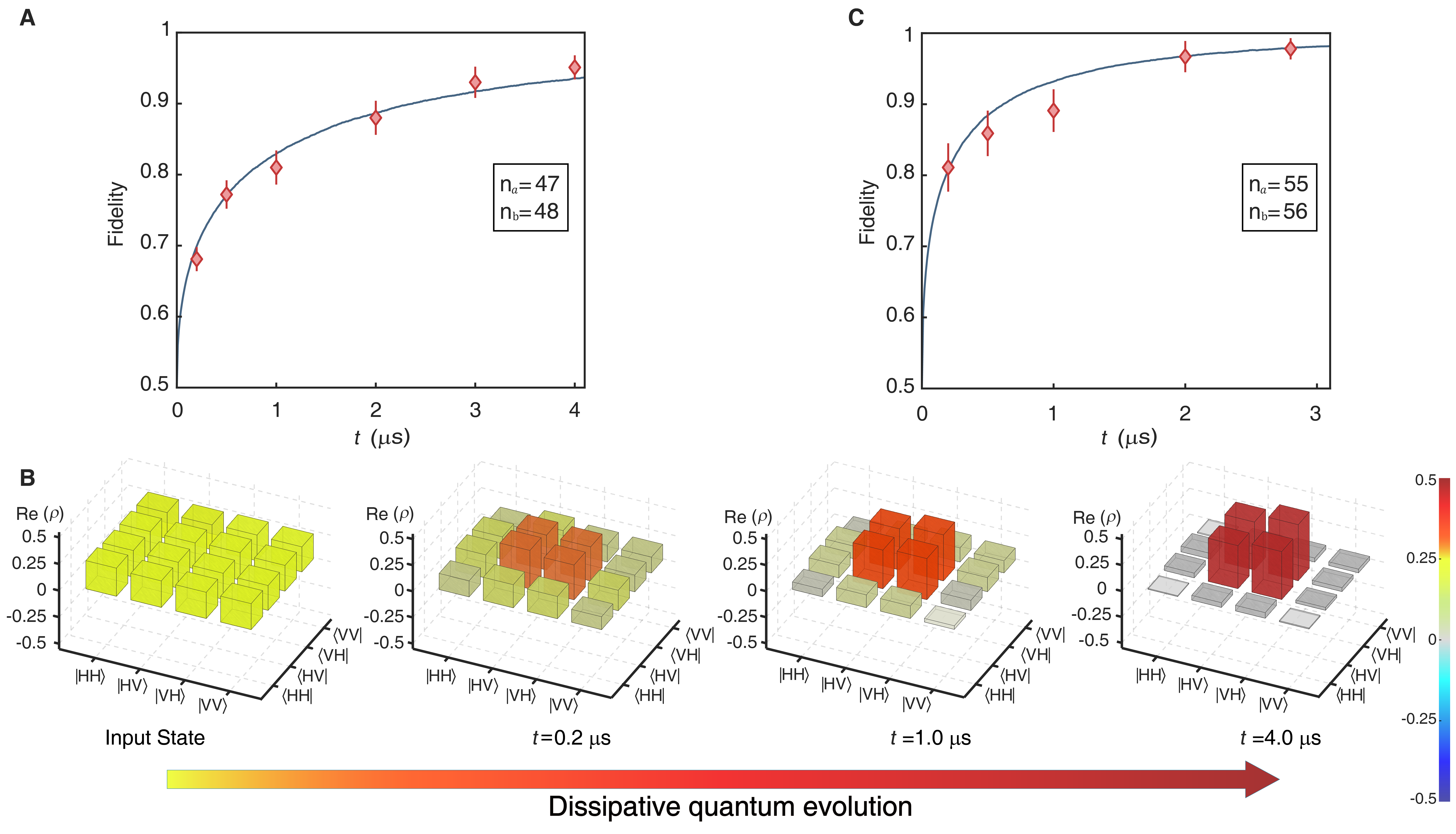}
  \caption{
	\textbf{
	Entanglement filter via dissipative quantum evolution.} \textbf{(A)} The measured state fidelity as a function of dissipative evolution time $t$ with $n_a=47, n_b=48$. 
	\textbf{(B)} The reconstructed density matrices for input state and output states with different evolution times.
	\textbf{(C)} Same as (A), but for $n_a=55, n_b=56$. 
	Error bars: $1\sigma$ standard deviation from photoelectric counting events.
	Solid curves: simulations based on the two-body decoherence
model.
	}
  \label{Fig:4}
\end{figure*}

Similar to $\chi$, the measured output-state fidelity depends on the quantum evolution time $t$ in the dissipative regime.
The observed temporal dynamics of output-state fidelity is shown in Fig.~4A, which agrees well with the result simulated by the two-body decoherence model~({\it 27\/}).
The undesired components in the density matrix $\rho$ are gradually suppressed with a longer dissipative evolution time (Fig.~4B).
After $\SI{4}{\micro\second}$ of evolution, the entanglement filter improves the state fidelity from $F_{in} = 0.495(7)$ to $F = 0.951(17)$.

The dissipative entanglement filter protocol is very robust to the principal quantum number $n$. For higher Rydberg states $n\sim55$, the larger interaction variation leads to a quicker dissipation of the unwanted photonic components (Fig.~3D) and consequently a faster improvement of the state fidelity with the evolution time $t$.
In Fig.~4C, an entanglement fidelity of $F=0.978(15)$ is achieved with $t = \SI{2.8}{ \mu\second}$.
In principle, the dissipative scheme can also be realized with very low $n$, by harnessing the resonant dipole-dipole interactions. 
Our simulation shows that, efficient entanglement filtering can be achieved even with $\ket{r_a}=\ket{19 D_{5/2},m_j=3/2}$ and $\ket{r_b}=\ket{20 P_{3/2},m_j=3/2}$ (see supplementary materials).
Combined with the blockade-based scheme at high $n$, the working range of our entanglement filter spans over a large spectrum of Rydberg states.

Although the transmission of the entanglement filter is determined by the efficiency of photon storage-and-retrieval processes, our protocol is intrinsically deterministic.
Currently, the measured storage efficiency $\eta_s = 0.24$ and read-out efficiency $\eta_r = 0.36$ are mainly limited by the finite optical depth ($\sim 3.5$) and can be further improved. By incorporating the atomic ensemble into an optical resonator, $\eta_s$ and $\eta_r$ can be increased to near unity and the high entanglement transmission could be achieved.
Moreover, the target state of our entanglement filter is not limited to $\ket{\Psi^+}$. In principle, our entanglement filter can extract any of the four Bell states from an arbitrary input states. For example, we also demonstrate the entanglement filtering for $\ket{\Psi^-}$ from an input state $\ket{H}_a\ket{V}_b=1/\sqrt{2}(\ket{\Psi^+}+\ket{\Psi^-})$, with an output state fidelity of $0.989(10)$ (see supplementary materials).

In summary, we report the realization of a novel entanglement filter enabled by Rydberg atoms. In contrast to previous demonstrations, our scheme features superior scalability, since it works without extra photonic resources, removes undesired states deterministically, and features the high entanglement fidelity.

Our entanglement filter opens new avenues for a number of novel quantum photonic applications and studies. 
First of all, by scaling up to an array of Rydberg ensembles, the efficient generation and manipulation of multi-photon entanglements, such as Dicke states and GHZ states, can be realized.
Moreover, our protocol is not based on the photon-photon interference, therefore, the qubits do not need to be indistinguishable. 
This unique feature allows an effective quantum control between photons with different temporal-spatial profiles and even with different colors, as long as they can be coupled to appropriate atomic transitions.
Last but not least, our entanglement filter also succeeds in the dissipative regime by exploiting the interaction-induced two-body decoherence.
The extension of such a scheme to an array of interacting Rydberg excitations could enable the dissipative preparation of long-range correlated states, such as Wigner crystal~({\it 28\/}), and the exploration of novel many-body quantum dynamics~({\it 29,30\/}) with interaction disorders that can be tuned by orders of magnitude.

\section*{Acknowledgments}
The authors thank Daiqin Su and Kuan Zhang for valuable discussions and Chenhao Du, Feng-Yuan Kuang and Yafen Cai for experimental assistance. This work was supported by the National Key Research and Development Program of China under Grants No.~2021YFA1402003, the National Natural Science Foundation of China (Grant No.~U21A6006, No.~12004127, and No.~12104173), and the Fundamental Research Funds for the Central Universities, HUST. Y.C. is supported by the National Natural Science Foundation of China (Grant No.~U2141237). T.S. is supported by National Key Research and Development Program of China (Grant No.~2017YFA0718304), and the National Natural Science Foundation of China (Grants No.~11974363, No.~12135018, and No.~12047503)

\section*{Author contributions}
All authors contributed substantially to this work.

\section*{Competing interests}
The authors declare no competing interests.

\section*{Data and materials availability}
Data that support the plots within this paper are available from the corresponding authors upon reasonable request. 

\clearpage


\section*{Supplementary material}

## I. EXPERIMENTAL DETAILS

In this section, we present the experimental details about the setup, EIT photon storage and the matter-light quantum state transfer.

$^{87}$Rb atoms are loaded into a magnetic-optical-trap (MOT) from the background vapor for 330 ms and a 50 ms long polarization-gradient-cooling (PGC) is performed to further lower the atomic temperature to $T = 10\,\mu$K. Then the atoms are loaded into a one-dimension optical dipole trap, formed by a linearly polarized 1012 nm laser beam along the y-axis, where the x-axis represents the vertical direction. The 1012 nm dipole trap beam has $\sim 0.9$ W power and waists of 10 $\mu$m and 50 $\mu$m, resulting in a trap depth of $U/h \sim 4$ MHz and trapping frequencies of $w_x/(2\pi) = 840$ Hz, $w_y/(2\pi) = 25$ Hz and $w_z/(2\pi) = 4$ kHz. The root-mean-squared (RMS) distributions of the atomic ensemble are $x_{RMS} \sim 15\,\mu$m, $y_{RMS} \sim 520\,\mu$m and $z_{RMS} \sim 3.5\,\mu$m. The 780 nm field propagates along the z-direction and is tightly focused onto the atoms with a waist of 6 $\mu$m, which defines the excitation region in the x-y plane. The excitation region in the z-direction is defined by length of the atomic sample $z_{RMS} \sim 3.5\,\mu$m. Before performing the entanglement filter protocol, a 8 G external magnetic field along z-axis is switch on and atoms are optically pumped to the $|g\rangle = |5S_{1/2}, F = 2, m_F = 2\rangle$ state. The atomic sample has an optical depth (OD) of $\sim 3.5$ for $|g\rangle - |e\rangle$ transition.

The 480 nm laser is from a second harmonic generator seeded by power-amplified 960 nm laser light. The 780 nm and 960 nm fields are generated from external cavity diode lasers, which are frequency-stabilized to an ultra-low-expansion cavity with a finesse of $\sim 20{,}000$. The linewidths of 780 nm and 960 nm lasers are $\sim 10$ kHz. An additional locking laser is employed to stabilize the relative phase between photons going through the upper and down paths. The wavelength of the locking laser is chosen to be 792 nm such that its AC-Stark shift and off-resonant scattering are negligible. The interference signal from the two 792 nm laser beams is detected to generate the feedback to a piezoelectric transducer (PZT) for stabilizing the optical path difference.

After every sample preparation, the 5 $\mu$s long entanglement filter protocol is executed 1,5000 times. Photons $a$ and $b$ are approximated by two 50 ns long weak coherent laser pulses separated by 100 ns. The horizontally (H) and vertically (V) polarized components are split into the down (D) and upper (U) paths by a polarization-beam-splitter (PBS). By applying 480 nm control laser fields $\Omega_a^{U,D}$ and $\Omega_b^{U,D}$, photons $a$ and $b$ are stored into the atomic ensemble. The photon storage is implemented via a ladder-type Rydberg electromagnetically induced transparency (EIT) scheme involving ground state $|g\rangle$, intermediate state $|e\rangle = |5P_{3/2}, F = 3, m_F = 3\rangle$ and Rydberg state $|r_a\rangle = |n_a D_{5/2}, J = 5/2, m_j = 5/2\rangle$ or $|r_b\rangle = |n_b D_{5/2}, J = 5/2, m_j = 5/2\rangle$. The Rydberg EIT has a bandwidth of $\Delta_{EIT} = 2\pi \times 4.4$ MHz. The 480 nm laser fields are right-hand circularly polarized and counter-propagate with the 780 nm photons. To match polarization of photonic qubits with $|g\rangle \leftrightarrow |e\rangle$ transition, quarter-wave plates (QWPs) are used to rotate the polarization of photons in the D (U) path from $|H\rangle$ ($|V\rangle$) to right-hand circular $|R\rangle$ before entering the atomic ensemble. After a storage time $t$ (200 ns for the blockade protocol and variable for the dissipative scheme), 480 nm read-out fields $\Omega_a^{U,D}$ and $\Omega_b^{U,D}$ are sequentially applied to convert the Rydberg excitations back into photons. The polarization of photons retrieved from D (U) are rotated back to $|H\rangle$ ($|V\rangle$) by QWPs. At the end of every experimental trial, the resonant 480 nm fields $\Omega_a^{U,D}$ and $\Omega_b^{U,D}$ are turned on again for 2.5 $\mu$s to clean any residue

Rydberg contaminants which might affect the Rydberg EIT storage in the next experimental trial. The optical dipole trap is switched off during the $5\,\mu s$ storage-and-retrieval experiment to avoid affecting the Rydberg EIT storage and turned back on for $60\,\mu s$ after every 5 experimental trials to keep the atoms confined.

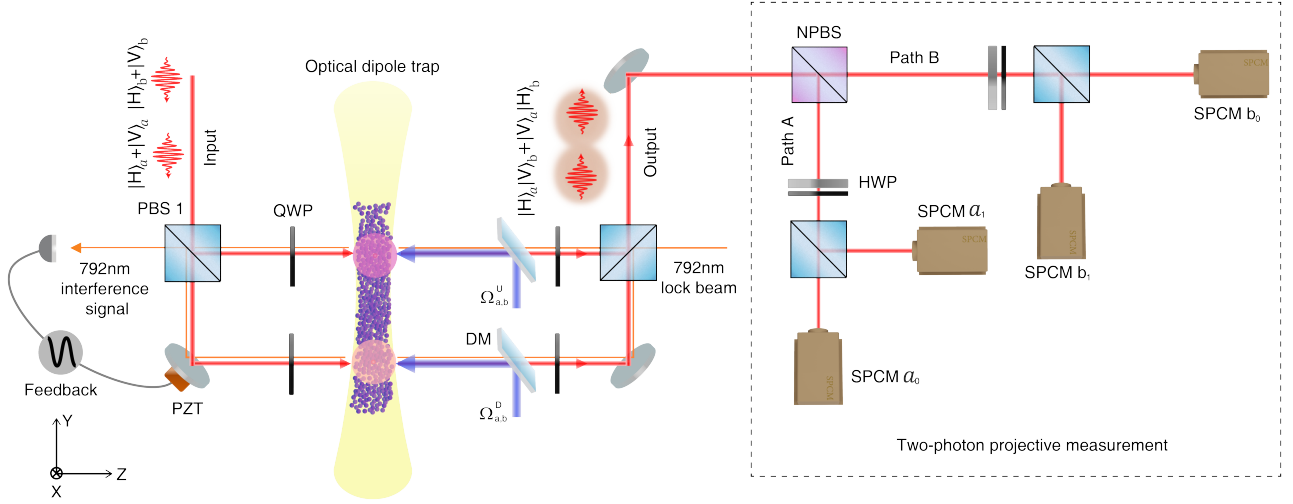

FIG. S1. **Schematic illustration of entanglement filter.** Two mesoscopic $^{87}$Rb atomic ensembles with optical depth of $\sim 3.5$ are confined by a one-dimensional optical dipole trap at $1012\,nm$. The $780\,nm$ photonic qubits $a$ and $b$ successively enter the system and their horizontally (H) and vertically (V) polarized components are separated into the down (D) and upper (U) paths by a polarization beam splitter (PBS). The qubits are stored as Rydberg excitations via the counter-propagating $480\,nm$ laser fields with the Rabi frequencies $\Omega_a^{U,D}$ and $\Omega_b^{U,D}$ under the electromagnetically induced transparency (EIT) condition. The $780\,nm$ and $480\,nm$ fields, with waists of $6\,\mu m$ and $15\,\mu m$, respectively, are tightly focused on the ensemble. Due to the strong Rydberg interactions, the storage of photons $a$ and $b$ into the same atomic ensemble, i.e., the transfer of $|H\rangle_a\,|H\rangle_b$ $(|V\rangle_a\,|V\rangle_b)$ into $|D\rangle_a\,|D\rangle_b$ $(|U\rangle_a\,|U\rangle_b)$, is suppressed. In the read-out stage, the $480\,nm$ laser fields $\Omega_a^{U,D}$ and $\Omega_b^{U,D}$ are sequentially applied to convert the Rydberg excitations back into photonic qubits. In order to store the $780\,nm$ photons using $|g\rangle - |e\rangle$ transition, the polarization of photons in the D (U) path is rotated from $|H\rangle\,(|V\rangle)$ to $|R\rangle$ before the EIT storage and converted back into $|H\rangle\,(|V\rangle)$ after the read-out. The output state of the entanglement filter is reconstructed via a polarization-projective measurement setup, which contains a non-polarization beam splitter (NPBS), half- and quarter-waveplates (HWPs and QWPs), PBSs, and single-photon counting modules (SPCMs). An additional $792\,nm$ locking laser beam is employed to stabilize the relative phase between photons going through the upper and down paths. The $792\,nm$ interference signal, which is measured at one of the ports in PBS 1, is used as the feedback to a piezoelectric transducer (PZT) to stabilize the phase difference between the optical paths. DM: dichroic mirror.

Here, we introduce how the storage efficiency $\eta_s = 24\%$ and retrieval efficiency $\eta_r = 36\%$ are measured. The $780\,nm$ field is detuned from the $|g\rangle - |e\rangle$ transition by $\Delta_{780} = -36\,MHz$ such that the absorption of $780\,nm$ photons without Rydberg EIT is negligible. The detuned storage efficiency, defined as the ratio of stored to input photons: $\eta_s^\Delta = \frac{N_{stor}}{N_{input}}$, is measured to be $\sim 2\%$. With $\eta_s^\Delta$, the retrieval efficiency $\eta_r = 36\%$ can be obtained from storage-and-retrieval efficiency $\eta_s^\Delta \times \eta_r = 0.73\%$ in the detuned storage ($\Delta_{780} = -36\,MHz$). The measured retrieval efficiency $\eta_r = 36\%$ should not be affected by the detuning in the storage process. Then the $780\,nm$ photon is tuned to resonant to optimize storage efficiency and we measure a resonant storage-and-retrieval efficiency of $\eta_s \times \eta_r = 8.5\%$, from which the resonant storage efficiency $\eta_s = 24\%$ in the entanglement filter (EF) experiment is obtained. The storage and retrieval efficiencies are mainly limited by the finite optical depth in the current experiment, and can be improved by incorporating the atomic ensemble with an optical resonator.

The polarization correlations of the output qubits can be obtained through the projective measurement shown in Fig S1. Two polarization-encoded qubits $a$ and $b$ are sequentially retrieved from the Rydberg ensembles, separated into paths A and B, and detected by two pairs of SPCMs. A non-polarization beam splitter (NPBS) is employed to simplify the experiment setup but it brings an additional optical loss of 50%. For example, we detect the photonic qubit $a$ at time $t_A$ in one of the outputs of the NPBS (path A), while the qubit $b$ at time $t_B$ is detected in another output port of the NPBS (path B). Alternatively, the two temporally-separated qubits can be routed into different paths with an electro-optic modulator (EOM), which has high optical transmission. The bases for polarization projective measurements in path A and B are determined by the waveplates (WPs) in front of the PBSs. Due to the collective photon emission in the EIT read-out, the retrieved photons have the same spatial profile as the $780\,nm$ input field and

can be conveniently coupled to the single-mode fiber, followed by the SPCMs $a_i$ and $b_j$ for $i, j = \{0, 1\}$. The collection angle of the phase-match emission is of $\sim 0.038$ rad. The correlation measurement is implemented by detecting the coincidences between SPCMs $a_i$ and $b_j$. To protect the SPCMs from the background lights, acousto-optic modulators (AOMs) are used before SPCMs as extra gates.

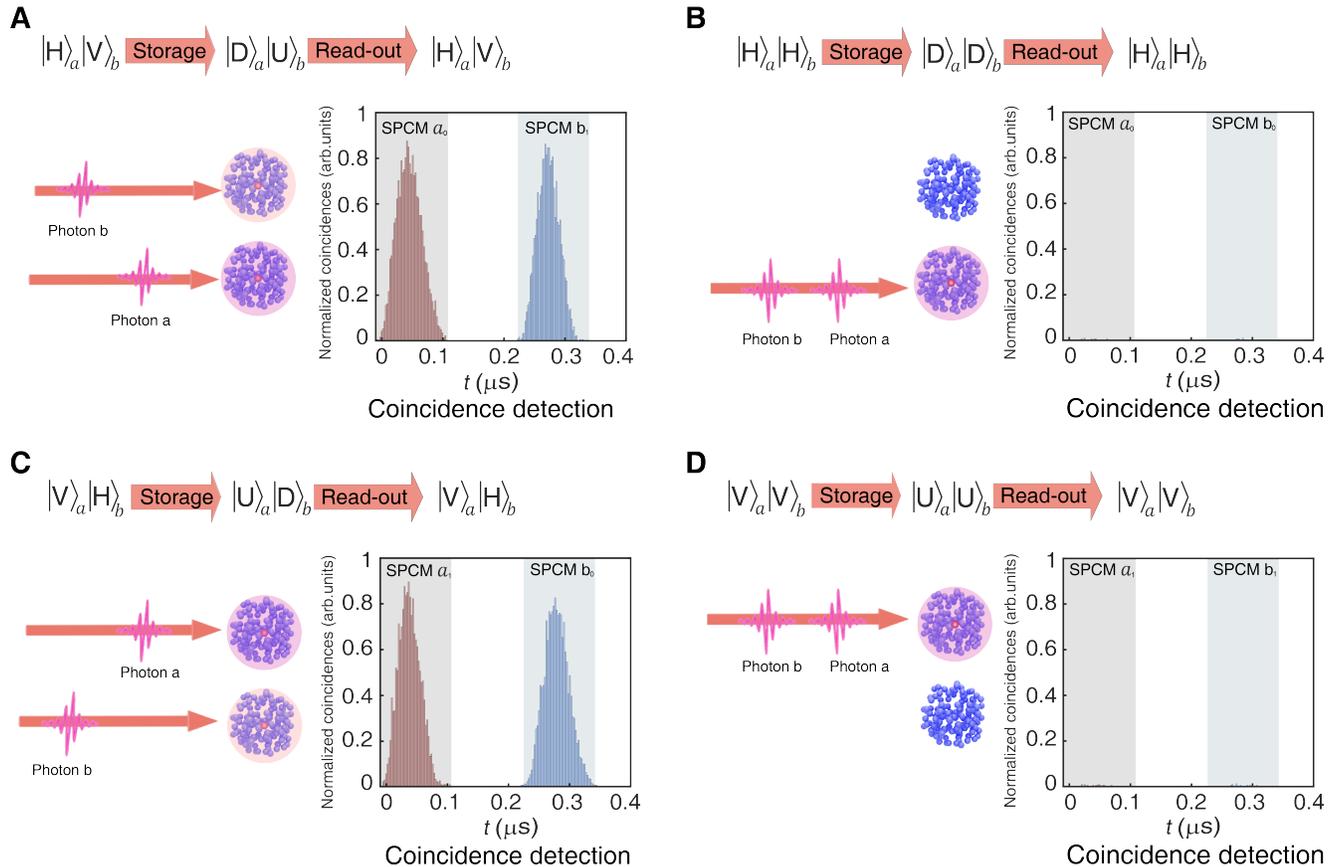

FIG. S2. **Time-resolved two-photon coincidences with different input states. (A and C)** Photons $a$ and $b$ are stored as Rydberg excitations in two individual ensembles $|D\rangle_a |U\rangle_b$ (A) and $|U\rangle_a |D\rangle_b$ (C), where the Rydberg interaction is negligible due to the $150 \, \mu m$ separation between the two atomic ensembles. Then the read-out light fields are sequentially applied to convert the Rydberg excitations to two retrieved photons centered at different time. The time-resolved two-photon coincidences between SPCM $a_i$ and $b_j$ are recorded. The temporal profiles of the retrieved photons are shown, with a full width at half maximum (FWHM) of about 45 ns. **(B and D)** Two photons entering the same ensemble $|D\rangle_a |D\rangle_b$ (B) and $|U\rangle_a |U\rangle_b$ (D) can not be simultaneously stored due to the strong Rydberg blockade effect. Therefore, the probability of retrieving photons $|H\rangle_a |H\rangle_b$ ($|V\rangle_a |V\rangle_b$) is highly suppressed, leading to almost vanishing correlated detection events in the readout.

## II. ENTANGLEMENT FILTER BASED ON RYDBERG BLOCKADE

### 1. Polarization-selective photon blockade

In this section, we present the supplementary information about the polarization-selective photon blockade illustrated in Fig. 2B of the main text and estimate the upper limit of output state fidelity for the blockade-based entanglement filter.

As illustrated in the main text, we use interactions between Rydberg states $|r_a\rangle = |76D_{5/2}, m_J = 5/2\rangle$ and $|r_b\rangle = |77D_{5/2}, m_J = 5/2\rangle$ to realize the polarization-selective photon blockade. Because of the strong Rydberg interaction, the storage of two photons in the same ensemble is suppressed, resulting in the normalized two-photon populations shown in Fig. 2, A and B. Figure S2 shows the time-resolved two-photon correlation measurements with different input qubits combinations. When photons $a$ and $b$ are routed into different ensembles, the Rydberg interactions are

negligible due to the large distance between two ensembles, so two photons can be stored and then retrieved at a later time (Fig. S2, A and C). However, the storage of two photons into double Rydberg excitations in the same ensemble is forbidden by the Rydberg interaction induced energy shifts. Note that only two-photon coincidences are recorded in Fig. S2, and therefore the correlated detection events for retrieving two photons from the same ensemble are strongly suppressed (Fig. S2, B and D).

Then we proceed to estimate the intrinsic limitation on the output state fidelity using the blockade-based EF protocol. The following results are acquired in the Rydberg-blockade regime but can easily be generalized to other cases. Considering a pure input state $|\varphi_{in}\rangle = \frac{1}{2}(|H\rangle_a |H\rangle_b + |H\rangle_a |V\rangle_b + |V\rangle_a |H\rangle_b + |V\rangle_a |V\rangle_b)$ for the EF operation, we can write the output state $|\varphi_{out}\rangle$ as:

$$|\varphi_{out}\rangle = \frac{1}{\sqrt{2 + P_{HH} + P_{VV}}} (\sqrt{P_{HH}} |H\rangle_a |H\rangle_b + |H\rangle_a |V\rangle_b + |V\rangle_a |H\rangle_b + \sqrt{P_{VV}} |V\rangle_a |V\rangle_b). \tag{1}$$

Here $P_{HH}(P_{VV})$ represents the residual probability of having output photons in the state $|H\rangle_a |H\rangle_b (|V\rangle_a |V\rangle_b)$ after the EF operation. Assuming the storage-and retrieval efficiencies and optical losses of ensemble D and U are identical, the suppression ratio can be expressed as $\chi = \frac{P_{HH} + P_{VV}}{P_{HV} + P_{VH}} = \frac{P_{HH}}{P_{HV}}$, where we have used $P_{VV} = P_{HH}$ and $P_{HV} = P_{VH}$. Consequently, Eq. 1 is simplified as

$$|\varphi_{out}\rangle = \frac{1}{\sqrt{2 + 2\chi}} (\sqrt{\chi} |H\rangle_a |H\rangle_b + |H\rangle_a |V\rangle_b + |V\rangle_a |H\rangle_b + \sqrt{\chi} |V\rangle_a |V\rangle_b). \tag{2}$$

and the theoretical upper limit of the output state fidelity is given by

$$F_{upp} = |\langle \Psi^+ |\varphi_{out}\rangle|^2 = \frac{1}{2} + \frac{1 - \chi}{2 + 2\chi}. \tag{3}$$

Therefore, in the ideal situation, the output state fidelity is only limited by the imperfection of the Rydberg blockade, characterized by the suppression ratio $\chi$. As a result of the strong interactions between Rydberg states $n_a = 76$ and $n_b = 77$, $\chi$ should be negligibly small. However, the background detection events, the multi-photon components from weak coherent light and the crosstalk due to the imperfection of polarization increase $\chi$ and lower the measured fidelity (thoroughly investigated in Supplementary Material Section II.2). To estimate the value of $\chi$ caused by the imperfect blockade effect, we measure the cross-correlation function $g_{ab}$ between photon a and b using only one ensemble, where the polarization crosstalk and multi-photon component in the weak coherent light do not play any role. The measured cross-correlation function $g_{ab} = 1.9(5) \times 10^{-3}$ is mainly contributed by the background detection events and the imperfect blockade. By analyzing the SPCMs background, we obtain the background-corrected cross-correlation function $g_{ab}^{cor} = 2(6) \times 10^{-4}$. Assuming the transmissions of photons through the path D and U are identical, the $\chi$ in Eq. 3 can be estimated by the cross-correlation function:

$$\chi = \frac{\rho_{HH} + \rho_{VV}}{\rho_{HV} + \rho_{VH}} = \frac{\frac{\langle N_{HH}\rangle}{\langle N_H\rangle \langle N_H\rangle} + \frac{\langle N_{VV}\rangle}{\langle N_V\rangle \langle N_V\rangle}}{\frac{\langle N_{HV}\rangle}{\langle N_H\rangle \langle N_V\rangle} + \frac{\langle N_{VH}\rangle}{\langle N_V\rangle \langle N_H\rangle}} = g_{ab}, \tag{4}$$

where $\langle N_H\rangle = \langle N_V\rangle$ is used. Therefore, we estimate that the upper limit for the output state fidelity is $F_{upp} = \frac{1}{2} + \frac{1 - g_{ab}^{(cor)}}{2 + 2g_{ab}^{(cor)}} = 0.9998(6)$.

Moreover, with the measured value $\chi = 0.0044(15)$ and its background-subtracted value $\chi_{BG} = 0.0025(15)$ shown in Fig. 2 of the main text, we attribute the remaining deviation of $\chi$ to the crosstalk due to the imperfection of polarization, which is estimated to be $\Delta\chi_{cross} = 0.0023(15)$.

### 2. Entangled state fidelity and error analysis

In this section, we characterize the quantum nature of our entanglement filter by performing photonic state tomography to reconstruct the density matrix of the output state and to measure its fidelity. Major error mechanisms and their contributions to the infidelity are thoroughly investigated.

The output state fidelity can be obtained with two-photon projective measurement on different Pauli bases: An arbitrary two-qubits state can be expanded by the identity and Pauli matrices as

$$\rho = \frac{1}{4} \sum_i \sum_j S_{ij} \sigma_i \otimes \sigma_j , \tag{5}$$

where $i, j = \{0, x, y, z\}$ correspond to the subscripts of different Pauli matrices:

$$\sigma_0 = \begin{pmatrix} 1 & 0 \\ 0 & 1 \end{pmatrix} \qquad \sigma_x = \begin{pmatrix} 0 & 1 \\ 1 & 0 \end{pmatrix} \qquad \sigma_y = \begin{pmatrix} 0 & -i \\ i & 0 \end{pmatrix} \qquad \sigma_z = \begin{pmatrix} 1 & 0 \\ 0 & -1 \end{pmatrix}. \tag{6}$$

Using Bell state $|\Psi^+\rangle = 1/\sqrt{2}(|H\rangle_a |V\rangle_b + |V\rangle_a |H\rangle_b)$ as the target state, the output state fidelity is

$$F_{\Psi^+} = \langle \Psi^+ | \rho | \Psi^+ \rangle = \frac{1}{4}(1 + S_{xx} + S_{yy} - S_{zz}) = \frac{1}{4}(1 + \langle \sigma_{xx} \rangle + \langle \sigma_{yy} \rangle - \langle \sigma_{zz} \rangle), \tag{7}$$

Where $\langle \sigma_{ij} \rangle = tr(\rho \sigma_{ij})$ is the expected value under the density matrix $\rho$. Therefore, the state fidelity can be obtained through projective measurements in three Pauli bases $\{\sigma_{xx}, \sigma_{yy}, \sigma_{zz}\}$. To reconstruct the output state density matrix $\rho$, we execute the state tomography via projective measurements on 9 different Pauli bases $\sigma_{ij}$ for $i, j = \{x, y, z\}$ with "2n detectors configuration (1, 2)". The maximum likelihood estimation technique is employed for the density matrix reconstruction. The reconstructed density matrix and its near-unity fidelity shown in Fig. 2 demonstrates the generation of near-perfect entangled photons and confirms the quantum nature of our Rydberg entanglement filter.

Next, we thoroughly analyze the major sources of error that reduce the state fidelity from the ideal value. We note that the following analysis is based on the entanglement filter experiment in the Rydberg blockade regime with $|r_a\rangle = |76D_{5/2}, m_j = 5/2\rangle$, $|r_b\rangle = |77D_{5/2}, m_j = 5/2\rangle$ (shown in Fig. 2 of the main text). However, the approaches used in the following analysis are universal and can be applied to general cases.

**Background detection events.** The background detection events mainly consist of the dark counts from the SPCMs, resulting in undesired coincidences in the correlation measurements. The expected value of $\langle \sigma_{ii} \rangle$ for $i = \{x, y, z\}$ is determined by

$$\langle \sigma_{ii} \rangle = P_{|i1\rangle} - P_{|i2\rangle} - P_{|i3\rangle} + P_{|i4\rangle}, \tag{8}$$

where $i = \{x, y, z\}$, $k = \{1, 2, 3, 4\}$, and $P_{|ik\rangle}$ represents the probability of detecting the k-th eigenstate of the $\sigma_{ii}$ basis. Taking $\sigma_{zz}$ basis as an example, we have $|z1\rangle = |H\rangle_a |H\rangle_b$, $|z2\rangle = |H\rangle_a |V\rangle_b$, $|z3\rangle = |V\rangle_a |H\rangle_b$, and $|z4\rangle = |V\rangle_a |V\rangle_b$. Then the expected value $\langle \sigma_{zz} \rangle$ is expressed as

$$\langle \sigma_{zz} \rangle = P_{|H\rangle_a |H\rangle_b} - P_{|H\rangle_a |V\rangle_b} - P_{|V\rangle_a |H\rangle_b} + P_{|V\rangle_a |V\rangle_b} = \frac{C_{00}(H,H) - C_{01}(H,V) - C_{10}(V,H) + C_{11}(V,V)}{C_{00}(H,H) + C_{01}(H,V) + C_{10}(V,H) + C_{11}(V,V)}. \tag{9}$$

Here $C_{ij}(\alpha_a, \alpha_b)$, with $i, j = 0, 1$ and $\alpha_a, \alpha_b = H, V$, is the coincidences between the SPCM $a_i$ and SPCM $b_j$. When a dark count occurs in one SPCM concurrently with a real photon detection event in another SPCM, a background-induced false coincidence is recorded (taking $C_{01}^{BG}(H,V)$ as an example):

$$C_{01}^{BG}(H,V) = p_{a_0}^{BG} p_{b_1} + p_{a_0} p_{b_1}^{BG}, \tag{10}$$

where $p_{a_0}^{BG}$ is the probability of background detection events in the SPCM $a_0$ and $p_{b_1}$ represents the probability of detecting a photon in the SPCM $b_1$. The coincidence between two background detection events can be neglected since $p_m^{BG} \ll p_n$, where $m, n = \{a_0, a_1, b_0, b_1\}$. The background-induced false coincidences $C_{00}^{BG}(H,H)$, $C_{10}^{BG}(V,H)$, $C_{11}^{BG}(V,V)$ have the same form, so the background-corrected value $\langle \sigma_{zz} \rangle^{cor}$ is given by

$$\langle \sigma_{zz} \rangle^{cor} = \frac{C_{00}(H,H) - C_{01}(H,V) - C_{10}(V,H) + C_{11}(V,V) - C_{00}^{BG}(H,H) + C_{01}^{BG}(H,V) + C_{10}^{BG}(V,H) - C_{11}^{BG}(V,V)}{C_{00}(H,H) + C_{01}(H,V) + C_{10}(V,H) + C_{11}(V,V) - C_{00}^{BG}(H,H) - C_{01}^{BG}(H,V) - C_{10}^{BG}(V,H) - C_{11}^{BG}(V,V)}. \tag{11}$$

Similarly, one can obtain the expression for $\langle \sigma_{xx} \rangle^{cor}$ and $\langle \sigma_{yy} \rangle^{cor}$. Therefore, the infidelity due to the background is

$$\Delta F_{BG} = \frac{1}{4}(\langle \sigma_{xx} \rangle^{cor} - \langle \sigma_{xx} \rangle + \langle \sigma_{yy} \rangle^{cor} - \langle \sigma_{yy} \rangle - \langle \sigma_{zz} \rangle^{cor} + \langle \sigma_{zz} \rangle) = 0.61(6)\%. \tag{12}$$

**Multi-photon components from weak coherent light.** As mentioned in the main text, we employ weak coherent laser pulses to approximate the single photons $a$ and $b$. The use of coherent fields as input only lowers the input and output state rates, but does not change the deterministic nature of our entanglement filter. The multi-photon components in the coherent field can contribute infidelity to the output photonic state. Here, we calculate an upper bound of the multi-photon-induced output state infidelity. An input coherent state after traveling through the PBS is written as

$$\sum_n \frac{c_n}{\sqrt{n!}} \left( \frac{\hat{a}_H^\dagger + \hat{a}_V^\dagger}{\sqrt{2}} \right)^n |0\rangle, \tag{13}$$

where $|0\rangle$ is the vacuum state and $a_H^\dagger(a_V^\dagger)$ is the creation operator for H (V)-polarized photons from the input 780 nm laser field. The coefficients of Poisson distribution with an average photon number $\langle n\rangle_{store}$ are depicted by $c_i^2$. For two incident pulses $a$ and $b$, the photon state after the PBS is

$$\sum_n \frac{c_n}{\sqrt{n!}}\left(\frac{a_{H_a}^\dagger + a_{V_a}^\dagger}{\sqrt{2}}\right)^n \otimes \sum_{n'} \frac{c_{n'}}{\sqrt{n'!}}\left(\frac{a_{H_b}^\dagger + a_{V_b}^\dagger}{\sqrt{2}}\right)^{n'}|0\rangle,\tag{14}$$

where the subscripts $a$ and $b$ denote photons from pulses $a$ and $b$, respectively. In the experiment, only two-photon coincidences are recorded, so the states with less than 2 photons have no contributions. Moreover, the higher-order terms with more than 3 photons are neglected since the average stored photon number $\langle n\rangle_{store} \ll 1$.

The H (V) polarized qubit state is sent to path D (U) by a PBS and stored in the corresponding atomic ensemble. For simplicity, we assume the storage-and-retrieval efficiencies for the unwanted multi-photon component are the same as that for single photon. In the strong Rydberg-blockade regime where multi-excitations in the same ensemble $|D_a D_b\rangle$ ($|U_a U_b\rangle$) are forbidden, the combined atom-photon state after the storage is given by

$$
\begin{aligned}
|\Psi_{stor}\rangle = \Big[ &\frac{c_1^2}{2}|D_a U_b\rangle + \frac{c_1^2}{2}|U_a D_b\rangle + \frac{c_1 c_2}{\sqrt{2}}\hat{a}_{H_b}^\dagger|D_a U_b\rangle + \frac{c_1 c_2}{\sqrt{2}}\hat{a}_{V_b}'^\dagger|U_a D_b\rangle \\
&+ \frac{c_1 c_2}{2\sqrt{2}}(\hat{a}'^\dagger_{H_a} + \hat{a}^\dagger_{V_b})|D_a U_b\rangle + \frac{c_1 c_2}{2\sqrt{2}}(\hat{a}^\dagger_{V_a} + \hat{a}'^\dagger_{H_b})|U_a D_b\rangle \Big]|0\rangle.
\end{aligned}\tag{15}
$$

The leading terms $\frac{c_1^2}{2}(|D_a U_b\rangle + |U_a D_b\rangle)|0\rangle$ represent the ideal case when the incident pulses contain only real single-photon components, while the remaining higher-order terms arise from the multi-photons components of the weak coherent states. Here, we use $\hat{a}'^\dagger_{H_{a/b}}$ and $\hat{a}'^\dagger_{V_{a/b}}$ as the creation operators for the unstored photons, to distinguish them from the incident and retrieved photons. In the experiment, only coincidences between the retrieved photons from Rydberg excitations are detected. Thus, we should consider the reduced density matrix $\rho_{Atom}$ by performing the partial trace in $|\Psi_{stor}\rangle\langle\Psi_{stor}|$ with respect to the subspace of the unstored photons:

$$
\begin{aligned}
\rho_{Atom} = &(\frac{c_1^4}{4} + \frac{3c_1^2 c_2^2}{4})|D_a U_b\rangle\langle D_a U_b| + (\frac{c_1^4}{4} + \frac{3c_1^2 c_2^2}{4})|U_a D_b\rangle\langle U_a D_b| + \\
&(\frac{c_1^4}{4} + \frac{c_1^2 c_2^2}{2})|D_a U_b\rangle\langle U_a D_b| + (\frac{c_1^4}{4} + \frac{c_1^2 c_2^2}{2})|U_a D_b\rangle\langle D_a U_b|.
\end{aligned}\tag{16}
$$

After the EIT storage, the 480 nm read-out fields are applied to convert the Rydberg excitations to retrieved photons described by

$$
\begin{aligned}
\rho_{Readout} = &(\frac{c_1^4}{4} + \frac{3c_1^2 c_2^2}{4})|H_a V_a\rangle\langle H_a V_a| + (\frac{c_1^4}{4} + \frac{3c_1^2 c_2^2}{4})|V_a H_b\rangle\langle V_a H_b| + \\
&(\frac{c_1^4}{4} + \frac{c_1^2 c_2^2}{2})|H_a V_b\rangle\langle V_a H_b| + (\frac{c_1^4}{4} + \frac{c_1^2 c_2^2}{2})|V_a V_b\rangle\langle V_a V_b|.
\end{aligned}\tag{17}
$$

Therefore, the output state fidelity is

$$F = \frac{\langle\Psi^+|\rho_{Readout}|\Psi^+\rangle}{tr(\rho_{Readout})} = 1 - \frac{1}{2}\frac{c_2^2}{c_1^2 + 3c_2^2}\tag{18}$$

With the average photon number of incident coherent light $\langle n\rangle = 0.10$ and the efficiency of storage $\eta_s = 0.24$, the average stored photon number is $\langle n\rangle_{store} = \langle n\rangle \times \eta_s = 0.024$. As a result, the infidelity resulting from the multi-photon components is given by $\Delta F_{mul} = 1 - F = \frac{1}{2}\frac{c_2^2}{c_1^2 + 3c_2^2} = 0.58\%$.

We emphasize that the analysis above is based on the assumption that the storage-and-retrieval efficiencies for the unwanted multi-photon components in Eq. 15 are the same as that for single photon. However, there are some mechanisms that reduce the efficiency in storing and retrieving these double excitations. For example, due to Rydberg interactions, the efficiency for coherently storing two photons to a single Rydberg excitation is lower than that of storing a single photon, leading to a reduced higher-order "error terms" population in Eq. 15 and smaller infidelity. Moreover, we ignored the dephasing of the collective Rydberg states induced by the transmitted photons ($\hat{a}'^\dagger_{H_{a/b}}$ ($\hat{a}'^\dagger_{V_{a/b}}$)) (3, 4), which induces dissipative emissions for higher-order "error terms" during read-out and alleviates

the contributed error. Therefore, the calculated $\Delta F_{mul} = 0.58\%$ only gives an upper bound of the multi-photon-induced infidelity. The detailed theoretical and experimental investigations of the multi-photon storage efficiency and atom-photon dephasing are interesting but beyond the scope of the current work.

When the stored photon number $\langle n \rangle_{store}$ increases, the expected output state fidelity will deviate from the Eq. 18 due to the contribution from three and more photons. However, similar analysis can still be performed by including higher-order multi-photon components. We emphasize that although weak coherent laser pulses are used to approximate single photons in our experiments, our entanglement filter protocol is intrinsically deterministic. The weak coherent laser pulses lead to low input state rate, but do not change the deterministic nature of our protocol and do not lower the transmission through the entanglement filter.

**Crosstalk due to polarization imperfections.** Due to imperfect optical polarization elements such as the wave-plates or PBSs, the projective measurements may not be performed exactly in the $\sigma_{ii}$ bases. Thus, polarization crosstalk occurs and induces undesired coincidences. Table S1 (left) shows the ideal projective measurements for state $|\Psi^+\rangle$, where the projections are done in the $\sigma_{ii}$ bases. With a small error arising from crosstalk $\xi \ll 1$, the corresponding results of the projective measurement are given in Table S1 (right). Table S1 clearly shows that a small fraction of the signal, which should have been projected to a particular basis, is incorrectly transferred to the orthogonal basis because of the polarization imperfection.

To estimate the infidelity caused by the polarization imperfection, we use the typical value of suppression ratio contributed by the crosstalk $\Delta\chi_{cross} = 0.0023(15)$ obtained in Section II.1, to extract the deviation of expected value $\langle \Delta\sigma_{zz} \rangle^{cross} = 0.34(30)\%$. Thus, the infidelity due to the imperfection of polarization is given by

$$\Delta F_{cross} = \frac{\langle \Delta\sigma_{xx} \rangle^{cross} + \langle \Delta\sigma_{yy} \rangle^{cross} + \langle \Delta\sigma_{zz} \rangle^{cross}}{4} = \frac{3 \langle \Delta\sigma_{zz} \rangle^{cross}}{4} = 0.25(23)\% \quad (19)$$

where $\langle \Delta\sigma_{xx} \rangle^{cross} = \langle \Delta\sigma_{yy} \rangle^{cross} = \langle \Delta\sigma_{zz} \rangle^{cross}$ is used because they have the same expression when $\xi \ll 1$.

| Ideally | $P_{|i1\rangle}$ | $P_{|i2\rangle}$ | $P_{|i3\rangle}$ | $P_{|i4\rangle}$ | Corsstalk | $P_{|i1\rangle}$ | $P_{|i2\rangle}$ | $P_{|i3\rangle}$ | $P_{|i4\rangle}$ |
|---------|---------|---------|---------|---------|-----------|---------|---------|---------|---------|
| XX | 0.5 | 0 | 0 | 0.5 | XX | $0.5 - 0.5\xi$ | $0.5\xi$ | $0.5\xi$ | $0.5 - 0.5\xi$ |
| YY | 0.5 | 0 | 0 | 0.5 | YY | $0.5 - 0.5\xi$ | $0.5\xi$ | $0.5\xi$ | $0.5 - 0.5\xi$ |
| ZZ | 0 | 0.5 | 0.5 | 0 | ZZ | $0.5\xi$ | $0.5 - 0.5\xi$ | $0.5 - 0.5\xi$ | $0.5\xi$ |

TABLE S1. **The result of projective measurements for Bell state $|\Psi^+\rangle$ with perfect polarization projection (left) and with a small polarization crosstalk (right).**

In summary, we have investigated major mechanisms responsible for reducing the state fidelity and estimated their individual contributions. Considering the infidelity of $0.61(6)\%$ from the background detection events, $\leq 0.58\%$ from the multi-photon component of weak coherent light, $0.25(23)\%$ from the crosstalk due to the imperfection of polarization, and the upper bound of $99.98(6)\%$ from imperfect double excitation suppression, we expect the state fidelity to be $98.5(2)\% \leq F_{exp} \leq 99.1(2)\%$, which is in good agreement with the measured value of $98.8(5)\%$. We emphasize that these error mechanisms are not fundamental to our entanglement filter protocol and can be alleviated with future technical efforts, such as using SPCMs with lower dark counts rate, employing real and high-quality single-photon source, improving polarization extinction ratio, and achieving stronger interaction with Rydberg states under Förster resonance.

## III. ENTANGLEMENT FILTER VIA DISSIPATIVE QUANTUM EVOLUTION

The van der Waals interactions between Rydberg states increase as $n^{11}$, leading to the strong blockade and efficient quantum control over photons using highly excited states. As shown in the previous section, we can use the polarization-selective photon blockade effect with high $n$ states to achieve an effective photonic entanglement filter. However, compared to low-lying states, high $n$ states suffer from various extra decoherence and losses, such as long-lived Rydberg contaminants, energy level shifts induced by residual electric fields and density-dependent dephasing.

Therefore, novel quantum photonic protocols with low $n$ states benefit from longer coherence times and hold the promise to tackle the above problems. For example, the Rydberg state with $n = 77$ has a short coherence time of $\sim 1.4\,\mu s$, while $n = 48$ state features a much longer $\sim 6\,\mu s$ coherence time in our experiment. In this section, we demonstrate a Rydberg-based quantum photonic operation with low principal quantum numbers $n \sim 48$, where the blockade effect is absent and the Rydberg-interaction induced two-body dissipation plays an essential role, as illustrated in the main text.

### 1. Detailed study of dissipative quantum evolution with Rydberg atoms

We first show theoretically how our entanglement filter extracts the Bell state $|H\rangle_a |V\rangle_b + |V\rangle_a |H\rangle_b$ from a classical input state by exploiting the dissipative quantum evolution. The suppression ratio $\chi$ and output state fidelity $F$ are simulated with the distribution of atomic ensemble given in section I.

**Photon storage.** We start from the photon storage process in the atomic ensembles. Under the resonant condition, the dynamics in the EIT storage process is governed by the Hamiltonian:

$$H_{stor} = g\sqrt{N} a A^\dagger + \Omega_C T_+ + H.c., \tag{20}$$

where $g$ is each atom's coupling strength between the 780 nm excitation field depicted by the annihilation operator $a$ (wavevector $\mathbf{k_1}$) and the atom's transition $|g\rangle \leftrightarrow |e\rangle$, $\Omega_C$ is the Rabi frequency of the 480 nm control light (wavevector $\mathbf{k_2}$), $N$ is the atom number, and the collective atomic transition operators $A$ and $T_- = T_+^\dagger$ are defined as

$$A = \frac{1}{\sqrt{N}} \sum_{j=1}^{N} e^{-i\mathbf{k_1} \cdot \mathbf{r}_j} \sigma_{ge}^j, \tag{21}$$

$$T_- = \sum_{j=1}^{N} e^{-i\mathbf{k_2} \cdot \mathbf{r}_j} \sigma_{er}^j. \tag{22}$$

Here, $\sigma_{\alpha\beta}^j = |\alpha\rangle_{jj} \langle\beta|$ is a quasi-spin operator for the $j$-th atom localizing at $\mathbf{r}_j$. We define an operator

$$D = a \cos\theta - C \sin\theta, \tag{23}$$

where

$$C = \frac{1}{\sqrt{N}} \sum_{j=1}^{N} e^{-i(\mathbf{k_1}+\mathbf{k_2}) \cdot \mathbf{r}_j} \sigma_{gr}^j \tag{24}$$

and the angle $\theta$ satisfies $\tan\theta = g\sqrt{N}/\Omega_C$. It can be proven (5) in the large $N$ limit with low excitations, i.e., most atoms are in the ground state $|g\rangle$, $[D, H_{stor}] = 0$ and $D$ is a bosonic annihilation operator satisfying $[D, D^\dagger] = 1$. As a result, the system has degenerate eigenstates with the eigenenergy 0 as

$$|\psi_{\{c_n\}}\rangle = \sum_n \frac{c_n}{\sqrt{n!}} D^{\dagger n} |G\rangle, \tag{25}$$

where $\{c_n\}$ is any given set of parameters and $|G\rangle$ is a state with all atoms in the ground state $|g\rangle$ and photons in the vacuum state.

Starting from the state $|G\rangle$, photon states can be stored in the atomic ensemble by adiabatically tuning $\theta$ from 0 to $+\pi/2$. In our experiment, the 780 nm excitation field is a weak coherent light with the average photon number $\bar{n} \ll 1$. Thus, after the first storage pulse, the two atomic ensembles (U and D) arrive at the following state

$$\begin{aligned}
|\psi_a\rangle &= \sum_n \frac{\bar{n}^{n/2}}{n! e^{\bar{n}/2}} \left( \frac{1}{\sqrt{2}} \left( C_a^{U\dagger} + C_a^{D\dagger} \right) \right)^n |G\rangle \\
&\approx \frac{1}{e^{\bar{n}/2}} |G\rangle + \frac{1}{\sqrt{2}} \sqrt{\frac{\bar{n}}{e^{\bar{n}}}} \left( C_a^{U\dagger} + C_a^{D\dagger} \right) |G\rangle
\end{aligned} \tag{26}$$

where the collective operator

$$C_\alpha^l = \frac{1}{\sqrt{N}} \sum_{j \in l} e^{-i(\mathbf{k}_1 + \mathbf{k}_2^\alpha) \cdot \mathbf{r}_j} \sigma_{gr_\alpha}^j \tag{27}$$

flips the quasi-spins in the $l$ ensemble from the Rydberg state $|r_\alpha\rangle$ to the ground state $|g\rangle$, with $\alpha = a, b$ and $l = \mathrm{U}, \mathrm{D}$. Here, we have expanded $|\psi_a\rangle$ up to one-excitation subspace because of the small $\bar{n}$. Applying the same procedure but with a 480 nm control field at different frequency, the second laser pulse is stored in the ensembles and the state becomes

$$|\psi_{ab}\rangle = \left[ \frac{1}{e^{\bar{n}/2}} + \frac{1}{\sqrt{2}} \sqrt{\frac{\bar{n}}{e^{\bar{n}}}} \left( C_b^{U\dagger} + C_b^{D\dagger} \right) \right] |\psi_a\rangle. \tag{28}$$

Here, the Rydberg interaction during the storage has been ignored since it is much smaller than the EIT storage bandwidth.

In the experiment, only two-photon coincidences are recorded. So in the following calculation, we will only consider the two-excitation state as the initial state and renormalize it as

$$|\psi(0)\rangle = \frac{1}{2} \left( C_b^{U\dagger} + C_b^{D\dagger} \right) \left( C_a^{U\dagger} + C_a^{D\dagger} \right) |G\rangle. \tag{29}$$

Then the atomic system evolves under the free atomic Hamiltonian and the Rydberg interaction. Therefore, apart from a general phase, the state at evolution time $t$ becomes

$$|\psi(t)\rangle = \frac{1}{2} \left( C_b^{U\dagger} C_a^{D\dagger} + C_b^{D\dagger} C_a^{U\dagger} \right) |G\rangle + \frac{1}{2N} \sum_{l=U,D} \sum_{jj' \in l} e^{i\phi_{jj'}(t)} \sigma_{r_b g}^{j'} \sigma_{r_a g}^j |G\rangle, \tag{30}$$

where

$$\phi_{jj'}(t) = \left( \mathbf{k}_1 + \mathbf{k}_2^b \right) \cdot \mathbf{r}_{j'} + \left( \mathbf{k}_1 + \mathbf{k}_2^a \right) \cdot \mathbf{r}_j - \frac{V_{jj'}}{\hbar} t. \tag{31}$$

In the next subsection, we will show how the photon emission from the second term in Eq. (30) decays with time $t$ because of the disordered phase $\phi_{jj'}(t)$.

**Read-out.** After the evolution time $t$, two 480 nm read-out fields with Rabi frequencies $\Omega_a^{U,D}$ and $\Omega_b^{U,D}$ and the corresponding momenta $\mathbf{k}_2^a$ and $\mathbf{k}_2^b$ are successively applied with a time delay $\tau$, inducing transitions between the states $|r\rangle_a / |r\rangle_b$ and $|e\rangle$. The decay rate $\gamma_e$ of the $|e\rangle$ state is comparable with $\Omega_a^{U,D}$ and $\Omega_b^{U,D}$, so the Rydberg atoms effectively decay to the ground states with a timescale $\sim 1/\gamma_e$ smaller than the read-out pulse duration, emitting photons $a$ and $b$. Using the input-output formalism where the input field has zero mean value, we can acquire the final state $|\Psi\rangle$ with two photons emitted and the atoms in the ground states as (6, 7, 8)

$$
\begin{aligned}
|\Psi\rangle &= \frac{1}{\mathcal{N}} \int \mathrm{d}\mathbf{k}_b \mathrm{d}\mathbf{k}_a e^{-ik_b \tau} \left[ \left( a_{\mathbf{k}_b H}^\dagger S_b^D a_{\mathbf{k}_a V}^\dagger S_a^U C_b^{D\dagger} C_a^{U\dagger} + a_{\mathbf{k}_b V}^\dagger S_b^U a_{\mathbf{k}_a H}^\dagger S_a^D C_b^{U\dagger} C_a^{D\dagger} \right) |G\rangle \right. \\
&\quad \left. + \frac{1}{N} a_{\mathbf{k}_b H}^\dagger S_b^D a_{\mathbf{k}_a H}^\dagger S_a^D \sum_{jj' \in D} e^{i\phi_{jj'}(t)} \sigma_{r_b g}^{j'} \sigma_{r_a g}^j |G\rangle + \frac{1}{N} a_{\mathbf{k}_b V}^\dagger S_b^U a_{\mathbf{k}_a V}^\dagger S_a^U \sum_{jj' \in U} e^{i\phi_{jj'}(t)} \sigma_{r_b g}^{j'} \sigma_{r_a g}^j |G\rangle \right] \\
&= \frac{1}{\mathcal{N}} \int \mathrm{d}\mathbf{k}_b \mathrm{d}\mathbf{k}_a e^{-ik_b \tau} \sum_{P_a, P_b = H, V} C_{\mathbf{k}_b P_b, \mathbf{k}_a P_a} a_{\mathbf{k}_b P_b}^\dagger a_{\mathbf{k}_a P_a}^\dagger |G\rangle, \tag{32}
\end{aligned}
$$

where $a_{\mathbf{k}\lambda}^\dagger$ is the photon's annihilation operator with momentum $\mathbf{k}$ and polarization $\lambda = H, V$, $\mathcal{N}$ is a normalization factor, $S_\alpha^l$ with $\alpha = a, b$ is defined similarly to $C_\alpha^l$

$$S_\alpha^l = \frac{1}{\sqrt{N}} \sum_{j \in l} e^{-i\mathbf{k}_2^\alpha \cdot \mathbf{r}_j} \sigma_{gr_\alpha}^j, \tag{33}$$

and the coefficients

$$C_{\mathbf{k}_b H, \mathbf{k}_a V} = \frac{1}{N^2} \sum_{j \in U, j' \in D} e^{i(\mathbf{k}_1 - \mathbf{k}_b) \cdot \mathbf{r}_{j'}} e^{i(\mathbf{k}_1 - \mathbf{k}_a) \cdot \mathbf{r}_j}, \tag{34}$$

$$C_{\mathbf{k}_b V, \mathbf{k}_a H} = \frac{1}{N^2} \sum_{j \in D, j' \in U} e^{i(\mathbf{k}_1 - \mathbf{k}_a) \cdot \mathbf{r}_{j'}} e^{i(\mathbf{k}_1 - \mathbf{k}_b) \cdot \mathbf{r}_j}, \tag{35}$$

$$C_{\mathbf{k}_b H, \mathbf{k}_a H}(t) = \frac{1}{N^2} \sum_{jj' \in D} e^{-i \frac{V_{jj'}}{\hbar} t} e^{i(\mathbf{k}_1 - \mathbf{k}_b) \cdot \mathbf{r}_{j'}} e^{-i(\mathbf{k}_1 - \mathbf{k}_a) \cdot \mathbf{r}_j}, \tag{36}$$

$$C_{\mathbf{k}_b V, \mathbf{k}_a V}(t) = \frac{1}{N^2} \sum_{jj' \in U} e^{-i \frac{V_{jj'}}{\hbar} t} e^{i(\mathbf{k}_1 - \mathbf{k}_b) \cdot \mathbf{r}_{j'}} e^{-i(\mathbf{k}_1 - \mathbf{k}_a) \cdot \mathbf{r}_j}. \tag{37}$$

Here, the factor $e^{-ik_b \tau}$ adds a read-out time delay $\tau$ to the $b$-photon, with $k_b$ the momentum magnitude along a given emission direction.

Because of the randomness of the atoms' positions, the coefficients $C_{\mathbf{k}_b H, \mathbf{k}_a V}$ and $C_{\mathbf{k}_b V, \mathbf{k}_a H} \approx 1$ for $\mathbf{k}_\alpha \approx \mathbf{k}_1$ ($\alpha = a, b$), while $C_{\mathbf{k}_b H, \mathbf{k}_a H}(t)$ and $C_{\mathbf{k}_b V, \mathbf{k}_a V}(t) \to 0$ for any $\mathbf{k}_a$ and $\mathbf{k}_b$ when $t$ is long enough so that the phases $e^{i \frac{V_{jj'}}{\hbar} t}$ get disordered. As a result, at the long evolution time limit, for two photons with momenta $\mathbf{k}_\alpha$ around $\mathbf{k}_1$, the final state is approximately

$$\frac{1}{\sqrt{2}} \left( |H\rangle_a |V\rangle_b + |V\rangle_a |H\rangle_b \right) = \frac{1}{\mathcal{N}'} \prod_{\alpha = a, b} \int_{\mathbf{k}_1} d\mathbf{k}_\alpha \left( a^\dagger_{\mathbf{k}_b V} a^\dagger_{\mathbf{k}_a H} + a^\dagger_{\mathbf{k}_b H} a^\dagger_{\mathbf{k}_a V} \right) e^{-ik_b \tau} |G\rangle, \tag{38}$$

where $\mathcal{N}'$ is a normalization factor and the integral is around the momentum $\mathbf{k}_1$. To see how this entangled state is generated, we define the suppression ratio $\chi(t) \equiv (P_{HH} + P_{VV}) / (P_{HV} + P_{VH})$. In the experiment only photons emitted along the direction $\mathbf{k}_1$ (with a small solid angle $\sim 0.005\,\text{rad}$) are collected and detected, so $\chi(t)$ can be approximated by

$$\chi \approx \frac{|C_{\mathbf{k}_1 H, \mathbf{k}_1 H}(t)|^2 + |C_{\mathbf{k}_1 V, \mathbf{k}_1 V}(t)|^2}{|C_{\mathbf{k}_1 H, \mathbf{k}_1 V}(t)|^2 + |C_{\mathbf{k}_1 V, \mathbf{k}_1 H}(t)|^2}$$
$$= \frac{1}{N^4} \left| \sum_{jj' \in U} e^{-i \frac{V_{jj'}}{\hbar} t} \right|^2. \tag{39}$$

Here, we have assumed in the last step that apart from the positions, the two atomic ensembles are identical.

Based on Eq. 39, the evolution of suppression ratio $\chi$ (Fig. 3, C and D) and output state fidelity $F$ (Fig. 4, A and C) with time can be numerically simulated with Monte-Carlo method. We first generate random positions configuration $\{r_1, ..r_i..., r_N\}$ for the $N \sim 440$ atoms in one ensemble that satisfy the geometry of the atomic sample mentioned in the Supplementary Material Section I. Next, we obtain the interaction-induced two-body phase shifts $V_{jj'} t/\hbar$ for all Rydberg pairs consisting of any two atoms separated by $R_{jj'}$ using the atomic-positions configuration $\{r_1, ..r_i..., r_N\}$. The distribution of the phase $\varphi_{jj'} \equiv (V_{jj'} t/\hbar) \, mod \, (2\pi)$ at different quantum evolution time $t$ is displayed in Fig. S3. At time $t = 0$, all atom pairs have the same phase 0, so the normalized probability $P_{Nor}$ is zero for $\varphi_{jj'} \neq 0$, as shown in Fig. S3A. The accumulation of random phases with longer evolution time $t$ is shown in Fig. S3B-E. The interaction leads to strongly disordered phases distribution and deteriorates the two-body coherence.

With the two-body phases distribution at hand, the suppression ratio $\chi$ and consequently the output state fidelity $F = \frac{1}{2} + \frac{1-\chi}{2+2\chi}$ can be obtained based on Eq. 39, by summing up the phase terms $e^{i \frac{V_{jj'}}{\hbar} t}$ for all Rydberg pairs. The evolution of $\chi$ strongly depends on the Rydberg interaction strength. To characterize the disorder of interaction, we define the standard deviation of interaction between Rydberg atomic pairs $dU = SD(V_{jj'}) = \frac{C_6}{SD(R^6_{jj'})}$. The $dU$ can be varied by changing the $C_6$ interaction coefficient with different Rydberg states $n_a$ and $n_b$. Besides the dissipative evolution using states $n_a = 47$ ($n_b = 48$) and $n_a = 55$ ($n_b = 56$) demonstrated in Fig. 3D, we also perform the experiment using states $n_a = 45$ ($n_b = 53$) (blue diamonds in Fig. S3). The $dU$ for states $n_a = 45, n_b = 53$ are two orders of magnitude weaker than that for states ($n_a = 55, n_b = 56$), leading to a slower decay of $\chi$ as a function of $t$. The simulated result remains very similar when the atom number $N$ is increased, showing that an atomic ensemble

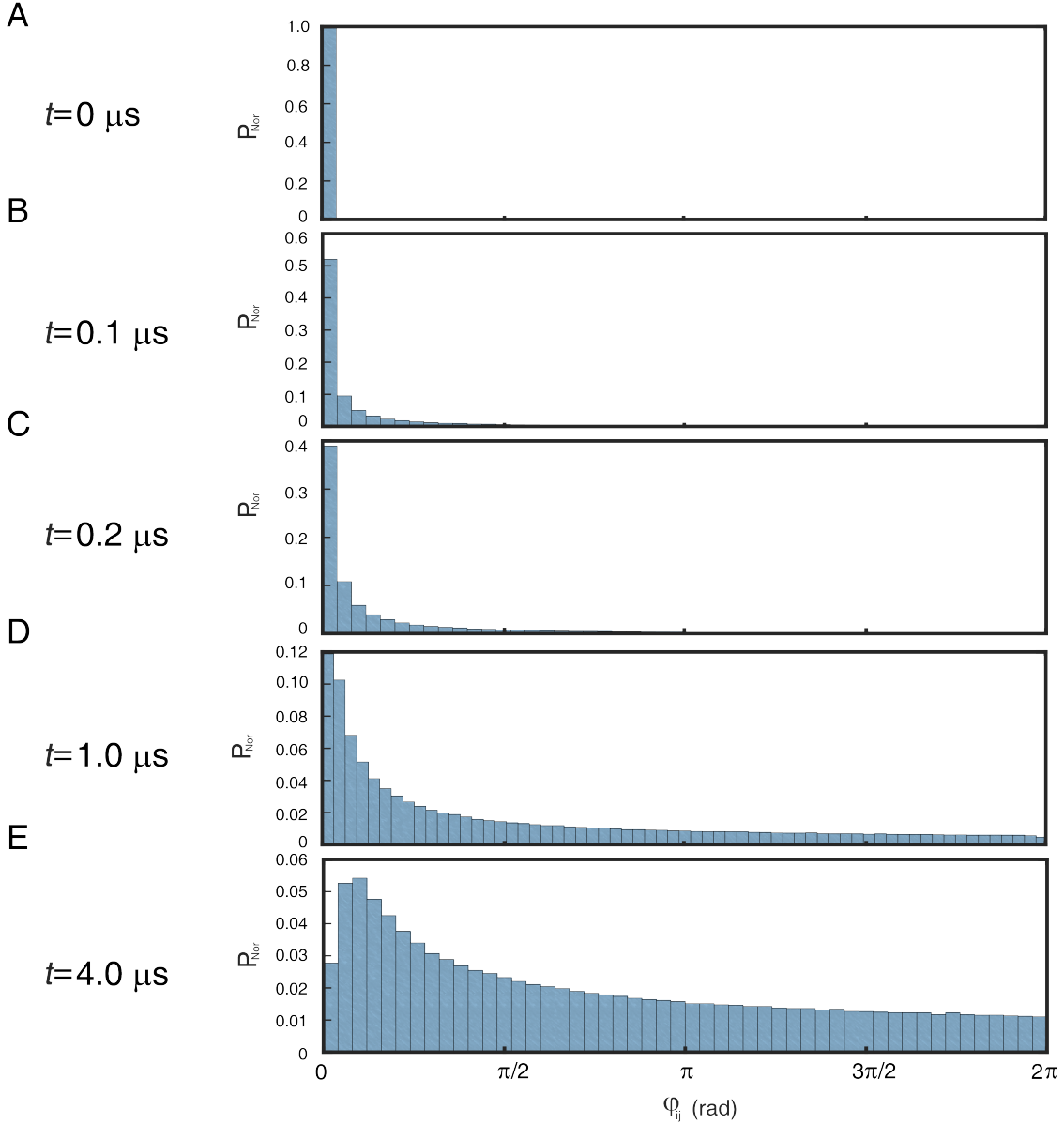

FIG. S3. **Evolution of the interaction-induced phase.** The normalized probability of the Rydberg interaction-induced phase $\varphi_{jj'} = (V_{jj'}t/\hbar) \, mod \, (2\pi)$ at different interaction time $t$ with states $n_a = 47, n_b = 48$.

with $N \sim 440$ can generate enough two-body decoherence for effective dissipative quantum state engineering. Note that apart from the atomic ensemble configuration and the Rydberg interaction $V_{jj'} = \frac{C_6}{R_{jj'}^6}$, our simulation contains no additional information or adjustable fitting parameters.

In the dissipative entanglement filter protocol, high output state fidelity is achieved after a storage time long enough to generate two-body dissipation. During this process, the single Rydberg excitation dephasing, induced mainly by the Doppler effect and density-dependent dephasing (*9, 10*), could lead to single-excitation dissipation during read-out. However, this single-excitation dissipation does not deteriorate the achieved output state fidelity.

To quantitatively investigate the influence of the single-excitation dephasing on output state fidelity, high principal quantum numbers $n_a = 76$ and $n_b = 77$ are used, so each atomic ensemble only has a single excitation due to the Rydberg blockade effect. The single-excitation dephasing time with $n \sim 77$ is about $1.4 \, \mu s$. The state fidelity with different storage time is shown in Table S2. We found that the output state fidelity stays around $F \sim 0.988$ when the storage time increases from $t = 200 \, \mathrm{ns}$ to $t = 1300 \, \mathrm{ns}$, which means that the single-excitation decoherence has

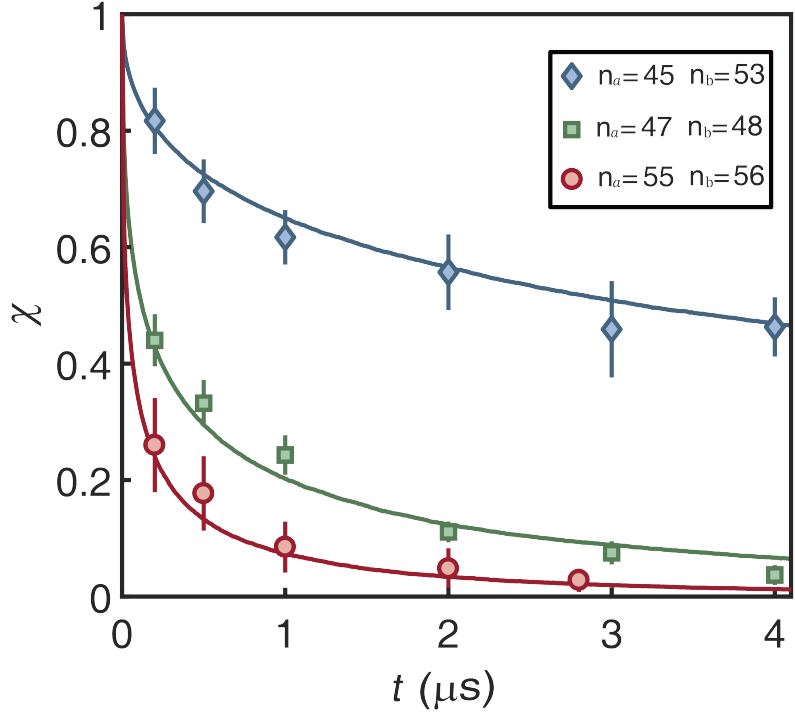

FIG. S4. **Dissipative quantum evolution with different interaction disorder.** The suppression ratio $\chi$ as a function of interaction time t with states $|r_a\rangle = |45D_{5/2}, J = 5/2, m_j = 5/2\rangle$, $|r_b\rangle = |53D_{5/2}, J = 5/2, m_j = 5/2\rangle$ (blue diamonds), $|r_a\rangle = |47D_{5/2}, J = 5/2, m_j = 5/2\rangle$, $|r_b\rangle = |48D_{5/2}, J = 5/2, m_j = 5/2\rangle$ (green squares), and $|r_a\rangle = |55D_{5/2}, J = 5/2, m_j = 5/2\rangle$, $|r_b\rangle = |56D_{5/2}, J = 5/2, m_j = 5/2\rangle$ (red circles). The solid lines are the corresponding theoretical simulations based on the two-body decoherence mechanism.

negligible effects on the output state fidelity.

| Storage time (ns) | 200 | 700 | 1300 |
|---|---|---|---|
| Fidelity | 0.988(5) | 0.985(18) | 0.987(27) |

TABLE S2. **Effect of single-excitation dephasing.** Output state fidelity as a function of the storage time with two Rydberg states $|r_a\rangle = |76D_{5/2}, J = 5/2, m_j = 5/2\rangle$, $|r_b\rangle = |77D_{5/2}, J = 5/2, m_j = 5/2\rangle$.

### 2. Dissipative entanglement filter using dipole–dipole interaction

Our dissipative entanglement filter protocol can, in principle, be applied to Rydberg states with very low principal numbers, by exploiting the dipole-dipole interaction $V_{dd} = C_3/R^3$ instead of the van der Waals interaction $V_{vdW} = C_6/R^6$. For example, a pair of Rydberg states with different parities, such as $|r_a\rangle = |nD_{5/2}, J = 5/2, m_j = 5/2\rangle$ and $|r_b\rangle = |(n + 1)P_{3/2}, J = 3/2, m_j = 3/2\rangle$, feature resonant dipole-dipole interaction. Compared with the EIT storage scheme illustrated in Fig. 1 of the main text, the photon storage in Rydberg $P$ state can be achieved with a control laser connecting $|e\rangle$-$nD_{5/2}, J = 5/2, m_j = 5/2\rangle$ transition and a MW field connecting the $|nD_{5/2}, J = 5/2, m_j = 5/2\rangle$ - $|(n + 1)P_{3/2}, J = 3/2, m_j = 3/2\rangle$ transition. Photon $a$ is first stored in Rydberg state $|nD_{5/2}, J = 5/2, m_j = 5/2\rangle$

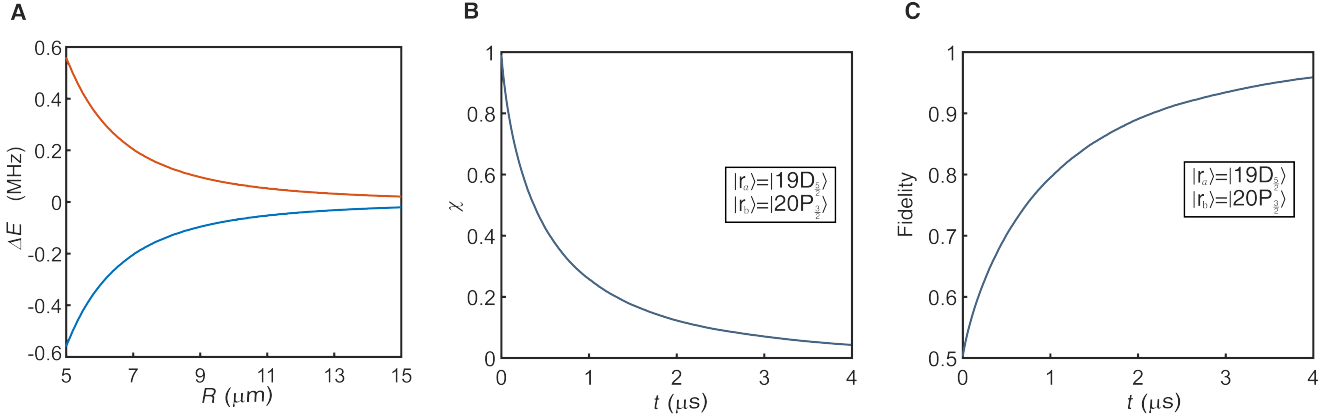

FIG. S5. **Dissipative quantum evolution using dipole-dipole interaction**. **(A)** The theoretical interaction-induced energy shift $\Delta E$ of pair state $|r_a r_b\rangle$ as a function of separation $R$ at $\theta = 0$ (atom aligned parallel to the quantization axis) calculated by diagonalizing the interaction Hamiltonian. The red line represent the symmetric state $|+\rangle = (|r_a r_b\rangle + |r_b r_a\rangle)/\sqrt{2}$ and the blue depict the anti-symmetric $|-\rangle = (|r_a r_b\rangle - |r_b r_a\rangle)/\sqrt{2}$. The two curves are fitted by the $C_3/R^3$ and give the coefficient of $C_3 = 70 \, \mathrm{MHz} \cdot \mu m^3$. **(B)** The theoretical simulation of $\chi$ as the function of interaction time $t$ with $|19D_{5/2}, J = 5/2, m_j = 5/2\rangle$ and $|20P_{3/2}, J = 3/2, m_j = 3/2\rangle$. **(C)** The theoretical simulation of output state fidelity $F$ as a function of interaction time $t$ with $|19D_{5/2}, J = 5/2, m_j = 5/2\rangle$ and $|20P_{3/2}, J = 3/2, m_j = 3/2\rangle$.

by the 480 nm control laser and then transferred to $|(n+1)P_{3/2}, J = 3/2, m_j = 3/2\rangle$ by a MW $\pi$−pulse. Photon $b$ is directly stored in Rydberg state $|nD_{5/2}, J = 5/2, m_j = 5/2\rangle$ with the 480 nm control laser. After the storage of both photons, the interaction-induced phase is accumulated between Rydberg states $|r_a\rangle = |nD_{5/2}, J = 5/2, m_j = 5/2\rangle$ and $|r_b\rangle = |(n+1)P_{3/2}, J = 3/2, m_j = 3/2\rangle$, resulting in dissipative quantum evolution. In the read-out process, photons $a$ and $b$ can be retrieved by reversing operations in the storage process.

The resonant dipole interaction leads to the symmetric and anti-symmetric states $|\pm\rangle = 1/\sqrt{2}(|r_a\rangle |r_b\rangle \pm |r_b\rangle |r_a\rangle)$, using the principal number $n = 19$. As shown in Fig. S5A, the interaction-induced energy shift $\Delta E$ of state $|+\rangle$ and $|-\rangle$ as a function of separation $R$ is calculated by diagonalizing the interaction Hamiltonian (*11*). The dipole-dipole interaction coefficient $C_3 \sim 70 \, \mathrm{MHz} \cdot \mu m^3$ can be extracted by fitting Fig. S5A with $C_3/R^3$. With the value of $C_3$, the evolution of suppression $\chi$ and output state fidelity $F$ are simulated and shown in Fig. S5, B and C. When the interaction time gradually increases to $4 \, \mu s$, the suppression ratio $\chi$ decreases from 1 to 0.04 and the output state fidelity is improved from 0.5 to 0.96. With the analysis above, we show that our entanglement filter can work efficiently and robustly over a very large Rydberg spectrum spanning from low-lying ($n \sim 19$) to a highly-excited ($n > 77$) states.

## IV. ENTANGLEMENT FILTERING FOR ARBITRARY STATES

We have shown that our entanglement filter can extract the target Bell state $|\Psi^+\rangle$ from a product state $\sim (\alpha |H\rangle_a + \beta |V\rangle_a)(\alpha |H\rangle_b + \beta |V\rangle_b)$ with an arbitrarily low initial entanglement fidelity. However, the target Bell state of our entanglement filter is not limited to $|\Psi^+\rangle$. For example, a proof-of-principle experiment that extract state $|\Psi^-\rangle$ from a input state of $|H\rangle_a |V\rangle_b = 1/\sqrt{2}(|\Psi^- + |\Psi^+\rangle)$ is performed in this section. Moreover, we show that, in principle, any of the four Bell states can be used as the target state and can be extracted from any input state containing the desired entanglement, by multiplexing the EF protocol and corresponding single-qubit rotations. This essential capability is important for many highly demanding applications in quantum photonics.

As a proof-of-principle demonstration of entanglement filtering beyond using $|\Psi^+\rangle$ as the target state, we perform an experiment that extracts the desired state $|\Psi^-\rangle$ from an input state of $|H\rangle_a |V\rangle_b = 1/\sqrt{2}(|\Psi^+\rangle + |\Psi^-\rangle)$, where the

four Bell states are defined as

$$|\Psi^+\rangle = \frac{1}{\sqrt{2}}(|H\rangle_a\,|V\rangle_b + |V\rangle_a\,|H\rangle_b)$$

$$|\Psi^-\rangle = \frac{1}{\sqrt{2}}(|H\rangle_a\,|V\rangle_b - |V\rangle_a\,|H\rangle_b)$$

$$|\Phi^+\rangle = \frac{1}{\sqrt{2}}(|H\rangle_a\,|H\rangle_b + |V\rangle_a\,|V\rangle_b)$$
(40)

$$|\Phi^-\rangle = \frac{1}{\sqrt{2}}(|H\rangle_a\,|H\rangle_b - |V\rangle_a\,|V\rangle_b).$$

Employing the states $n_a = 76$ and $n_b = 77$, the experiment is performed in the Rydberg blockade regime. Before sending the input state $|H\rangle_a\,|V\rangle_b$ to the EF, a Hadamard gate is performed on both qubits $a$ and $b$ with a HWP. The Hadamard gate changes state $|\Psi^+\rangle$ to $|\Phi^-\rangle$ and keeps the $|\Psi^-\rangle$ component invariant. The following EF operation blocks the $|\Phi^-\rangle$, while transmitting the target Bell state $|\Psi^-\rangle$. Figure S6, A and B, show the reconstructed density matrices for the input and the output states, respectively. The measured input state fidelity is $F_{\Psi^-} = \langle\Psi^-|\,\rho_{in}\,|\Psi^-\rangle = 0.495(7)$ and the entanglement filter improves it to $F_{\Psi^-} = \langle\Psi^-|\,\rho_{out}\,|\Psi^-\rangle = 0.989(10)$. The near-unity state fidelity demonstrates again our entanglement filter's capability of extracting desired entangled state from a input state with low initial fidelity.

One of the important application of EF is entanglement distillation, which requires the capability of improving the state fidelity from an input state with fidelity lowered by either "spin-flip" or "phase-flip" error. In this context, the EF operation that extracts $|\Psi^+\rangle$ from $1/2(|H\rangle_a\,|H\rangle_b + |H\rangle_a\,|V\rangle_b + |V\rangle_a\,|H\rangle_b + |V\rangle_a\,|V\rangle_b) = 1/\sqrt{2}(|\Psi^+\rangle + |\Phi^+\rangle)$ can be considered as distilling desired entanglement from a input state with "spin-flip error", while the EF that distills $|\Psi^-\rangle$ from $|H\rangle_a\,|V\rangle_b = 1/\sqrt{2}(|\Psi^+\rangle + |\Psi^-\rangle)$ corresponds to entanglement distillation from an input state with "phase-flip error" .

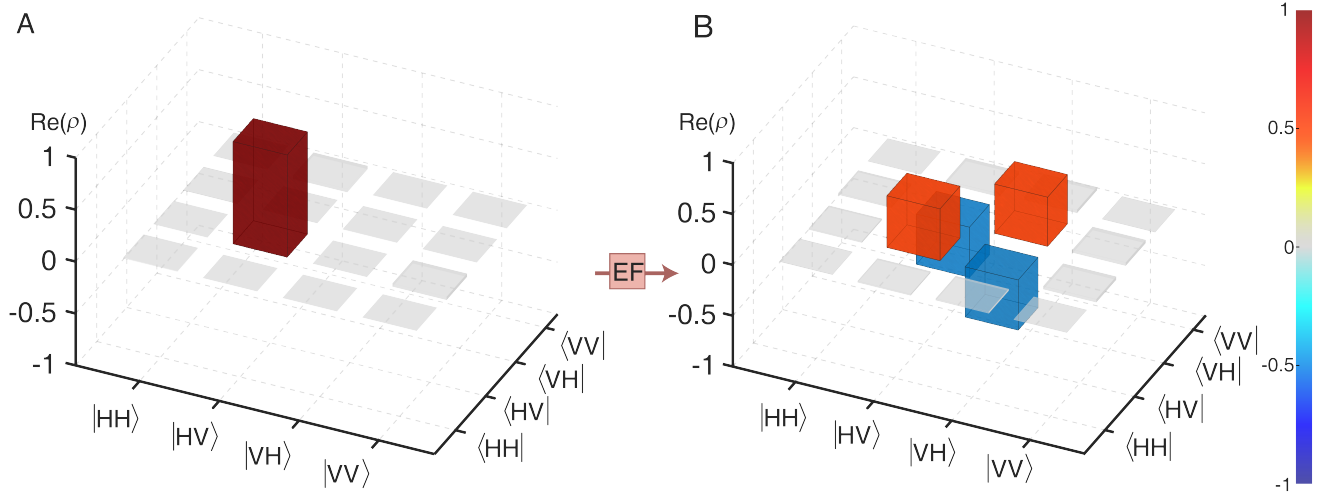

FIG. S6. **Quantum state filtering using the $|\Psi^-\rangle$ as the target state. (A and B)** The reconstructed density matrix of the input state $|H\rangle_a\,|V\rangle_b$ (A) and output state $|\Psi^-\rangle = 1/\sqrt{2}(|H\rangle_a\,|V\rangle_a - |V\rangle_b\,|H\rangle_b)$ (B).

Moreover, any of the four Bell states can, in principle, be used as the target state and can be extracted from arbitrary mixed states. Without loss of generality, we assume the density matrix of the initial state $\hat{\rho}_{initial}$ is diagonalized in the Bell bases:

$$\hat{\rho}_{initial} = b_1\,|\Psi^+\rangle\,\langle\Psi^+| + b_2\,|\Psi^-\rangle\,\langle\Psi^-| + b_3\,|\Phi^+\rangle\,\langle\Phi^+| + b_4\,|\Phi^-\rangle\,\langle\Phi^-|,$$
(41)

where $b_j$ with $j = 1, 2, 3, 4$ are the corresponding probabilities. We first consider the target output state as $|\Psi^+\rangle$. After passing through the entanglement filter, the $|H\rangle_a\,|H\rangle_b$ and $|V\rangle_a\,|V\rangle_b$ components in $\hat{\rho}_{initial}$ are eliminated due to the Rydberg blockade effect or dissipative quantum evolution, leading to an output state of

$$\hat{\rho}_{out1} = b_1\,|\Psi^+\rangle\,\langle\Psi^+| + b_2\,|\Psi^-\rangle\,\langle\Psi^-|.$$
(42)

Before sending $\hat{\rho}_{out1}$ to the second entanglement filter, a Pauli-Z rotation is performed on the $V$-polarized component in photonic qubit $b$, such that states $|\Psi^+\rangle$ and $|\Psi^-\rangle$ are exchanged: $\hat{\rho}_{out1} \rightarrow b_2\,|\Psi^+\rangle\,\langle\Psi^+| + b_1\,|\Psi^-\rangle\,\langle\Psi^-|$. This rotation

can be accomplished by using an EOM is to add a $\pi$ phase to the $V$-polarized component in photonic qubit $b$. Next, a Hadamard gate is executed on both qubits $a$ and $b$ with a half-wave plate, which changes the polarization $|H\rangle$ to $(|H\rangle + |V\rangle)/\sqrt{2}$ and $|V\rangle$ to $(|H\rangle - |V\rangle)/\sqrt{2}$. After above rotations, the input state $\hat{\rho}_{in2}$ for the second entanglement filter is

$$\hat{\rho}_{in2} = b_2 |\Phi^-\rangle \langle \Phi^-| + b_1 |\Psi^-\rangle \langle \Psi^-|. \tag{43}$$

Then, state $|\Phi^-\rangle$ is removed by the second entanglement filter, while state $|\Psi^-\rangle$ is transmitted. Therefore, the output state after the second filtering process is given by $\hat{\rho}_{out2} = b_1 |\Psi^-\rangle \langle \Psi^-|$. Finally, we perform the Pauli-Z rotation again to obtain the desired entangled state $\hat{\rho}_{final} = b_1 |\Psi^+\rangle \langle \Psi^+|$. Similar operations can be applied, when other Bell states are chosen as the target state. Figure S7 summarizes the required operation combinations of entanglement filters and Pauli rotations to extract any of the four Bell states from $\hat{\rho}_{initial}$. In principle, Bell states can also be extracted using the EF operations based on linear-optical approaches. However, compared with our intrinsically deterministic scheme, the linear-optical EF operations are probabilistic and require extra ancillary quantum resources, resulting in limited scalability.

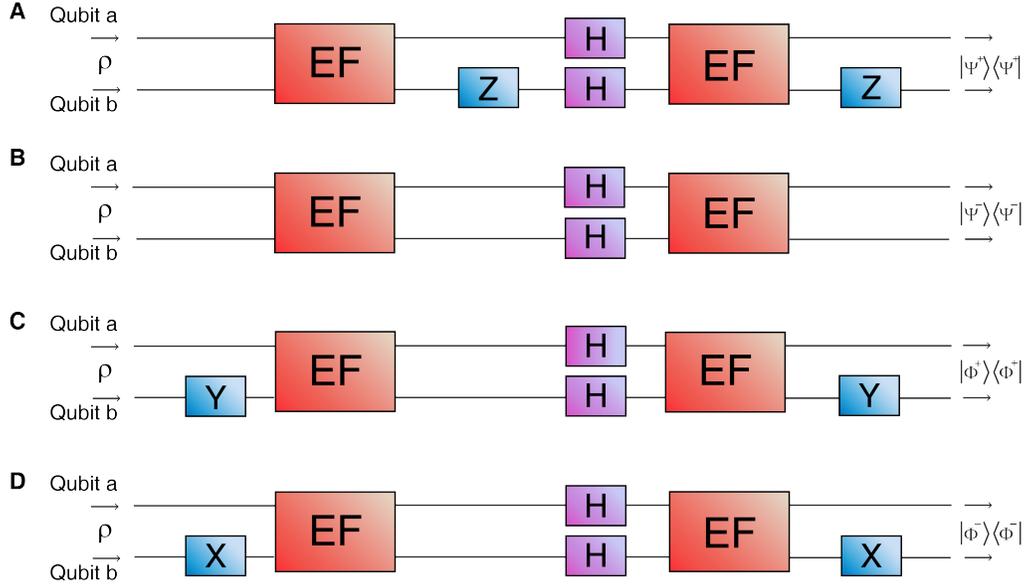

FIG. S7.   **Quantum state filtering for any of four Bell states from arbitrary mixed states. (A–D)** The quantum circuit diagram for multiplexing the EF operation to extract the chosen Bell states. All the input states are mixed states: $\hat{\rho}_{initial} = b_1 |\Psi^+\rangle \langle \Psi^+| + b_2 |\Psi^-\rangle \langle \Psi^-| + b_3 |\Phi^+\rangle \langle \Phi^+| + b_4 |\Phi^-\rangle \langle \Phi^-|$. The Pauli-X rotation X maps $|\Psi^\pm\rangle \leftrightarrow |\Phi^\pm\rangle$, the Pauli-Y rotation Y maps $|\Psi^\pm\rangle \leftrightarrow |\Phi^\mp\rangle$, and the Pauli-Z rotation Z maps $|\Psi^+\rangle \leftrightarrow |\Psi^-\rangle$, $|\Phi^+\rangle \leftrightarrow |\Phi^-\rangle$. The Hadamard gates H are executed by a HWP to realize a bilateral $\pi/2$ operation, which transforms $|\Psi^+\rangle \leftrightarrow |\Phi^-\rangle$).

We note that in some linear-optical EF literature, the two-qubits polarization configuration for the target state of the entanglement filter is defined differently. Devices that block photon pairs with different (*12, 13*) or the same (*14, 15*) polarizations are sometimes called entanglement filter/splitter (EF/ES), which can be described by the $EF/ES$ matrices in the $\{|H\rangle_a |H\rangle_b, |H\rangle_a |V\rangle_b, |V\rangle_a |H\rangle_b, |V\rangle_a |V\rangle_b\}$ basis as

$$EF = \begin{pmatrix} 1 & 0 & 0 & 0 \\ 0 & 0 & 0 & 0 \\ 0 & 0 & 0 & 0 \\ 0 & 0 & 0 & 1 \end{pmatrix} \qquad ES = \begin{pmatrix} 0 & 0 & 0 & 0 \\ 0 & 1 & 0 & 0 \\ 0 & 0 & 1 & 0 \\ 0 & 0 & 0 & 0 \end{pmatrix}. \tag{44}$$

Here, we do not distinguish the EF and ES operations, since our protocol features the capability of extracting any of the four Bell states from an input of arbitrary mixed state.


\end{document}